\documentclass[11pt,a4paper,oneside]{article}

\usepackage[a4paper]{geometry}
\usepackage{graphicx}
\usepackage{amsmath}
\usepackage{natbib}
\usepackage[normalem]{ulem} 

\usepackage{rotating}
\newlength\lengtha 
\newlength\lengthb 
\newlength\lengthc 

\usepackage{adjustbox}

\usepackage{booktabs} 
\usepackage{graphicx}
\usepackage{enumerate}

\usepackage{url} 

\usepackage{xr} 

\usepackage{tikz} 
\newcommand{\chck}{\checkmark}

\def\T{{ \mathsf{\scriptscriptstyle T} }}
\usepackage[ruled]{algorithm2e}

\usepackage{mathtools} 

\usepackage{enumerate} 

\newcounter{Algocount}
\newcounter{tempFigure}

\usepackage{array}
\newcolumntype{L}{>{\arraybackslash}m{.92\textwidth}}

\usepackage{bbm} 

\newcommand{\E}{\mathbb{E}}

\usepackage{amsfonts} 
\usepackage{eurosym}

\usepackage{epsf}
\usepackage{graphicx}
\usepackage{rotating}

\usepackage[latin1]{inputenc}

\newcommand{\beq}{\begin{equation}}
\newcommand{\eeq}{\end{equation}}

\newcommand{\rit}{{\rm I\!R}}

\newcommand{\ttf}[1]{{\ttfamily{#1}}}

\newcommand{\ee}[1]{\mathrm{e}^{#1}}

\newcommand{\tr}[1]{\mathrm{Tr}\left({#1}\right)}

\newcommand{\kron}{\otimes}
\newcommand{\diag}{\mathrm{diag}}

\newcommand{\N}[2]{{\cal N}\left(#1,#2\right)}

\newcommand{\G}[2]{{\cal G }\left(#1,#2\right)}

\newcommand{\ul}[1]{{\underline{#1}}}

\renewcommand{\Vec}[1]{\mathrm{vec}{(#1)}}

\DeclareMathAlphabet\mathbfcal{OMS}{cmsy}{b}{n}

\newcommand{\floor}[1]{\left\lfloor #1 \right\rfloor}

\usepackage[latin1]{inputenc}

\begin{document}

\title{
  Exogenous time-varying covariates in double additive cure
  survival model with application to fertility
}

\bibliographystyle{chicago}



\date{January 31, 2023}

  \author{Philippe Lambert\footnote{
    Institut de Math\'ematique,
    Universit\'e de Li\`ege, Belgium. Email: p.lambert@uliege.be}
    \footnote{Institut de Statistique, Biostatistique et Sciences Actuarielles
      (ISBA), Universit\'e catholique de Louvain, Belgium.}
  and Michaela Kreyenfeld\footnote{Hertie School Berlin, Germany.}}



\maketitle

\begin{abstract}
  Extended cure survival models enable to separate covariates that
  affect the probability of an event (or {\em long-term} survival)
  from those only affecting the event {\em timing} (or {\em
    short-term} survival).  We propose to generalize the bounded
  cumulative hazard model to handle additive terms for time-varying
  (exogenous) covariates jointly impacting long- and short-term
  survival. The selection of the penalty parameters is a challenge in
  that framework. A fast algorithm based on Laplace approximations in
  Bayesian P-spline models is proposed. The methodology is motivated
  by fertility studies where women's characteristics such as the
  employment status and the income (to cite a few) can vary in a
  non-trivial and frequent way during the individual follow-up.
  The method is furthermore illustrated by drawing on register data from
  the German Pension Fund which enabled us to study how women's
  time-varying earnings relate to first birth transitions. 
\end{abstract}

\noindent%

\section{Introduction} \label{Intro:Sec}

Proportional hazards models are used extensively to analyze
time-to-event data and their association to covariates. They enable to
summarize group differences using risk ratios assumed constant over
time. Cure survival models \citep{Bo1949,BeGa1952} explicitly
acknowledge that a proportion of the studied population will never
experience the event of interest whatever the duration of the
follow-up. This can be revealed or confirmed with the inspection of
the estimated survival functions (such as Kaplan-Meier curves) found
to reach a plateau at a non-zero level for large values of the
follow-up time.  We will focus here on the {\em promotion time} (cure)
survival model, also named the {\em bounded cumulative hazard model}
\citep{Yakovlev:1996,Tsodikov:1998,Chen:1999vqba}.  Let
$\mathbf{v}=(\mathbf{z},\mathbf{x})$ where $\mathbf{z}$ denote a
$p-$vector of categorical covariates, and $\mathbf{x}$ a $J-$vector of
quantitative covariates (with $\mathbf{0}$ generically used to refer
to their reference values).  If $S_p(t|\mathbf{v})$ is the conditional
survival function for subjects (including cured individuals) sharing
these characteristics, then
\begin{align}
  S_p(t|\mathbf{v}) = \exp\{-\vartheta(\mathbf{v}) F(t)\}
  \label{BasicPromotionTime:Eq}
\end{align}
where $t>0$, $\vartheta(\mathbf{\mathbf{v}})>0$ and $F(t)$ is a cumulative
distribution function such that $F(0)=0$ and $F(T)=1$ with $T$
denoting the minimal survival time after which a subject can be
declared cured.
The proportion of cured subjects in the sub-population defined by
$\mathbf{v}$ is
$$\pi(\mathbf{v}) = S_p(T|\mathbf{v})
= \exp\{-\vartheta(\mathbf{v})F(T))\} =
\exp\{-\vartheta(\mathbf{v})\}>0.$$
Let $f(t)=dF(t)/dt$. When
$\vartheta(\mathbf{v}) =\exp\{\eta_\vartheta(\mathbf{v})\}$ with
$\eta_\vartheta(0)=\beta_0$, (\ref{BasicPromotionTime:Eq}) corresponds
to a proportional hazards (PH) model with baseline hazard
$\mathrm{e}^{\beta0}f(t)$ and a cumulative hazard bounded by
$\vartheta(\mathbf{v})$. The dynamics in the hazard function, governed
by $f(t)$, is not affected by covariates: it ensures the constant
hazard ratio characterizing the PH model.  This is a crucial
assumption that is not always properly assessed with potential
consequences on the quality of the conclusion derived from the research.
Even when the estimated survival curves for the compared
groups do not cross and the parallelism of the logarithm of the
underlying cumulative hazards is not challenged during a follow-up
interrupted by right censoring, the PH hypothesis might be violated
further out in time. Indeed, the survival probabilities could
increasingly diverge or become similar beyond the largest observation
time, or even converge in the longer term.
In the latter case, a lower risk reported in the
treatment group from right-censored data would only indicate a
delayed event rather than a long-term treatment gain. It motivated 
the {\em extended promotion time} model \citep{Bremhorst:2016a,Bremhorst:2016b}
\begin{align}
  S_p(t|\mathbf{v},\tilde{\mathbf{v}}) = \exp\{-\vartheta(\mathbf{v}) F(t|\tilde{\mathbf{v}})\}.
  \label{ExtendedPromotionTime:Eq}
\end{align}
with a dynamics in the (population) hazard function changing with the
covariates in $\tilde{\mathbf{v}}$.
An accelerated failure time (AFT) or a proportional hazards (PH) model
could be considered further to describe the dependence of $F$ on
$\tilde{\mathbf{v}}$.

The inclusion of time-varying covariates (TVCs) is challenging in that
framework. It was studied in the mixture cure model formulation by
\citet{Dirick2019} with TVCs restricted to the
conditional survival model for non-cured subjects with, therefore,
only constant covariates entering the logistic regression submodel for
the cure probability. A more general formulation was proposed
by \citet{LambertBremhorst2020} in the framework of the extended promotion
time model with categorical TVCs affecting not only the event timing
for non-cured subjects, but also entering the regression model for the 
cure probability. However, the follow-up duration after each change of
covariate had to be long enough for identification purposes in the cure
probability submodel.

In the current paper, a reformulation of the extended promotion time
model allowing an unlimited number of changes in categorical or
continuous TVCs is proposed and studied. Additive terms for nonlinear
effects of (constant or time-varying) quantitative covariates can also be
considered jointly in the long- and short-term survival submodels.

The plan of the paper is as follows. In Section
\ref{ExtendedPromotionTime:Sec}, we propose a detailed recall of the
extended promotion time model. The inclusion of additive terms using
P-splines will also be discussed. The methodological core
of the paper is in Section \ref{TVcure:Sec} with a novel proposal for
the inclusion of categorical or quantitative time-varying covariates
in a cure survival model. Algorithms to explore the joint posterior of
the model parameters and to compute their posterior mode (MAP) are
described in Section \ref{Inference:Sec}. A strategy for selecting
penalty parameters tuning the smoothness of the unknown functionals in
the long- and short-term survival sub-models is proposed with a simple
to implement and fast converging algorithm. The merits of this
proposal are evaluated by means of an extensive simulation study in
Section \ref{SimulationStudy:Sec}. The methodology is illustrated in
Section \ref{Application:Sec} with the analysis of pension register
data and of the association between women's time-varying
earnings and fertility transitions in Germany. We conclude the paper
with a discussion in Section \ref{Discussion:Sec}.

\section{The extended promotion time model} \label{ExtendedPromotionTime:Sec}
The use of covariates to alter the dynamics in the nonparametric
baseline hazard of the promotion time model was first explored by 
\citet{Bremhorst:2016a}, see (\ref{ExtendedPromotionTime:Eq}), with
a log-linear model for $\vartheta(\mathbf{v})$ and PH model for
$F(t|\tilde{\mathbf{v}})$. More specifically, consider the following
formulation for the latter expression, 
$F(t|\tilde{\mathbf{v}}) = 1-S_0(t)^{\exp(\eta_F(\tilde{\mathbf{v}}))}$
where $S_0(t)=1-F(t|\mathbf{0})$ is a baseline survival function and $\eta_F(\cdot)$ is a (possibly
non-linear) function of the covariates with an identification constraint, for example
$\eta_F(\mathbf{0})=0$. Then, the population cumulative hazard and hazard
functions associated to (\ref{ExtendedPromotionTime:Eq}) become, respectively,
\begin{align}
  H_p(t|\mathbf{v},\tilde{\mathbf{v}}) & = \vartheta(\mathbf{v}) F(t|\tilde{\mathbf{v}})
  = \exp\{\eta_\vartheta(\mathbf{v})\} \left(1-S_0(t)^{\exp(\eta_F(\tilde{\mathbf{v}}))}\right) ~;
                                         \label{ExtendedPromotionTime:PopHazard:Eq} \\
  h_p(t|\mathbf{v},\tilde{\mathbf{v}}) &= \vartheta(\mathbf{v}) f(t|\tilde{\mathbf{v}})
 = \mathrm{e}^{\eta_\vartheta(\mathbf{v})+\eta_F(\tilde{\mathbf{v}})}
                            f_0(t)S_0(t)^{\exp(\eta_F(\tilde{\mathbf{v}}))-1} ~,
  \label{ExtendedPromotionTime:Pophazard:Eq}
\end{align}
where $f_0(t)=-dS_0(t)/dt$. 
Identification issues can be solved
provided that the follow-up is sufficiently long, even in the
challenging case where some covariates are common to $\mathbf{v}$ and
$\tilde{\mathbf{v}}$, see \citet{LambertBremhorst:2019} for more details.
For fixed given values $\tilde{\mathbf{v}}$ of the short-term survival
covariates, (\ref{ExtendedPromotionTime:Pophazard:Eq}) defines a
proportional hazards model with a cured fraction
$\exp(-\vartheta(\mathbf{v}))$ and non time-varying hazard ratios for
contrasts corresponding to different values of $\mathbf{v}$ as
$
h_p(t|\mathbf{v}_2,\tilde{\mathbf{v}}) /
h_p(t|\mathbf{v}_1,\tilde{\mathbf{v}})
= \vartheta(\mathbf{v}_2) / \vartheta(\mathbf{v}_1)
$.
This is
not true anymore when hazards are compared for different values of
$\tilde{\mathbf{v}}$, as
$
h_p(t|\mathbf{v},\tilde{\mathbf{v}}_2) /
h_p(t|\mathbf{v},\tilde{\mathbf{v}}_1)
= \tilde{\vartheta}(\tilde{\mathbf{v}}_2) /
\tilde{\vartheta}(\tilde{\mathbf{v}}_1)
S_0(t)^{\tilde{\vartheta}(\tilde{\mathbf{v}}_2) - \tilde{\vartheta}(\tilde{\mathbf{v}}_1)}
$
(with
$\tilde{\vartheta}(\tilde{\mathbf{v}})=\mathrm{e}^{\eta_F(\tilde{\mathbf{v}})}$)
changes over time. 

\citet{Bremhorst:2016a} considered a linear combination of
B-splines to specify $h_0(t)=f_0(t)/S_0(t)$. Here, a
flexible form based on P-splines \citep{EilersMarx:1996} is
preferred for
$f_0(\cdot)$,
\begin{align}
  f_0(t)= \frac{\exp\left(\sum_kb_k(t)\phi_k\right)}{
  \int_0^T \exp\left(\sum_kb_k(u)\phi_k\right)du}
  \mathbbm{1}_{\{0\leq t \leq T\}}
  \label{f0:Bsplines:Eq}
\end{align}
where
$\{b_k(\cdot):k=1,\ldots,K\}$ denotes a large B-splines basis
associated to equidistant knots on $(0,T)$ and
$\pmb{\phi}=(\phi_k)_{k=1}^K$ is a vector of spline parameters with
$\phi_{\floor{K/2}}=0$ (for identification purposes).
It is directly connected to the reference population hazard,
$h_p(t|\mathbf{0},\mathbf{0})=\mathrm{e}^{\beta_0}f_0(t)$, with
$\beta_0$ governing the total risk exposure and $f_0(t)$ its
distribution over time.
Smoothness will
be forced on $f_0(\cdot)$ by penalizing changes in the spline
coefficients, see Section \ref{Inference:Sec}.
The possible nonlinear effects of continuous covariates (such as age
or earnings on the probability of pregnancy and its timing in a
fertility study) will be modelled using additive forms.  Assume that
$n$ independent units are observed with data
${\cal D}=\cup_{i=1}^n{\cal D}_i$ where
${\cal D}_i=\{t_i,\delta_i,\mathbf{v}_i,\tilde{\mathbf{v}}_i\}$,
$\delta_i$ is the event indicator for the follow-up time $t_i$ and
$\mathbf{v}_i,\tilde{\mathbf{v}}_i$ the long- and short-term survival
covariates for unit $i$. We add flexibility to the extended promotion
time model by considering nonlinear forms to describe the
effects of quantitative covariates on $\eta_\vartheta(\mathbf{v})$ and
$\eta_F(\tilde{\mathbf{v}})$,
\begin{align}
\big(\eta_\vartheta(\mathbf{v}_{i})\big)_{i=1}^n
&= \left(\beta_{0}+\sum_{k=1}^p\beta_k z_{ik}
+\sum_{j=1}^{J} {f_{j}}(x_{ij})\right)_{i=1}^n
=\mathbf{Z} \pmb{\beta}
+\sum_{j=1}^{J}\mathbf{f}_j~, \label{EtaShortTerm:Eq}\\
\big(\eta_F(\tilde{\mathbf{v}}_{i})\big)_{i=1}^n
&= \left(\sum_{k=1}^{\tilde{p}} \gamma_k \tilde{z}_{ik}
+\sum_{j=1}^{\tilde{J}} {{\tilde{f}_{j}}}(\tilde{x}_{ij})  \right)_{i=1}^n
= \tilde{\mathbf{Z}} \pmb{\gamma}
+\sum_{j=1}^{\tilde{J}}\tilde{\mathbf{f}}_j~,  \label{EtaLongTerm:Eq}
\end{align}
where $f_{j}(\cdot)$ and $\tilde{f}_{j}(\cdot)$ denote smooth
additive terms quantifying the effect of the associated quantitative
covariate on long- and short-term survival, respectively,
$\mathbf{f}_j=\big(f_j(x_{ij})\big)_{i=1}^n$ and
$\tilde{\mathbf{f}}_j=\big(\tilde{f}_j(\tilde{x}_{ij})\big)_{i=1}^n$ their
values over units stacked in vectors, $\mathbf{Z}$ the $n\times (1+p)$ design matrix
with a column of 1's for the intercept and one column per additional
categorical covariate,
similarly for the $n\times \tilde{p}$ design matrix
$\tilde{\mathbf{Z}}$ (but without the column of 1's given the absence
of an intercept).
Now consider a basis of $(L+1)$ cubic B-splines
$\{s^*_{j\ell}(\cdot)\}_{\ell=1}^{L+1}$ associated to equally spaced knots
on the range $(x_j^{\min},x_j^{\max})$ of values for $x_j$.
They are recentered for identification purposes in the additive model using
$s_{j\ell}(\cdot)= s^*_{j\ell}(\cdot) - \tfrac{1}{x_j^{\max}-x_j^{\min}}\int_{x_j^{\min}}^{x_j^{\max}}s^*_{j\ell}(u)du~(\ell=1,\ldots,L)$.
Similarly for the covariates associated to short-term survival,
yielding recentered B-splines denoted by $\tilde{s}_{j\ell}(\cdot)$.
Then, the additive terms in the conditional long-term and
short-term survival sub-models can be approximated using linear combinations of such (recentered)
B-splines,
$\mathbf{f}_j=\left(\sum_{\ell=1}^L s_{j\ell}(x_{ij})\theta_{\ell j}\right)_{i=1}^n=\mathbf{S}_j\pmb{\theta}_j$, 
$\tilde{\mathbf{f}}_j=\left(\sum_{\ell=1}^L \tilde{s}_{j\ell}(\tilde{x}_{ij})\tilde{\theta}_{\ell j}\right)_{i=1}^n=\tilde{\mathbf{S}}_j\tilde{\pmb{\theta}}_j,$
where $[\mathbf{S}_j]_{i\ell}=s_{j\ell}(x_{ij})$, $[\tilde{\mathbf{S}}_j]_{i\ell}=\tilde{s}_{j\ell}(\tilde{x}_{ij})$,
$\big(\pmb{\theta}_j\big)_\ell=\theta_{\ell j}$
and $\big(\tilde{\pmb{\theta}}_j\big)_\ell=\tilde{\theta}_{\ell j}$.
Hence, using vectorial notations, the expressions for the conditional
long-term and short-term linear predictors in (\ref{EtaShortTerm:Eq})
and (\ref{EtaLongTerm:Eq})
can be rewritten as
$\big(\eta_{\vartheta i}=\eta_{\vartheta}(\mathbf{v}_{i})\big)_{i=1}^n
={\mathbfcal X} \pmb{\psi}$,
$\big(\eta_{Fi} = \eta_F(\tilde{\mathbf{v}}_{i})\big)_{i=1}^n
={\tilde{\mathbfcal X}} \tilde{\pmb{\psi}}$
with design matrices ${{\mathbfcal X}}=[\mathbf{Z},\mathbf{S}_1,\ldots,\mathbf{S}_J]
=[\mathbf{Z},\mathbfcal{S}]\in\rit^{n\times q}$,
${{\tilde{\mathbfcal X}}}=[\tilde{\mathbf{Z}},\tilde{\mathbf{S}}_1,\ldots,\tilde{\mathbf{S}}_J]
=[\tilde{\mathbf{Z}},\tilde{\mathbfcal{S}}]\in\rit^{n\times \tilde{q}}$;
matrices of spline parameters (with one column per additive term)
$\mathbf{\Theta}=[\pmb{\theta}_1,\ldots,\pmb{\theta}_J]$ in $\rit^{L\times J}$,
$\tilde{\mathbf{\Theta}}=[\tilde{\pmb{\theta}}_1,\ldots,\tilde{\pmb{\theta}}_{\tilde{J}}]$ in $\rit^{L\times \tilde{J}}$;
vectors of (stacked) regression parameters
$\pmb{\psi} = \begin{pmatrix}\pmb{\beta}, \Vec{\mathbf{\Theta}}\end{pmatrix}$ in $\rit^q$,
$\tilde{\pmb{\psi}} = \begin{pmatrix}\pmb{\gamma}, \Vec{\tilde{\mathbf{\Theta}}}\end{pmatrix}$
in $\rit^{\tilde{q}}$, where $q=(1+p+JL)$ and $q=(\tilde{p}+\tilde{J}L)$.
Note that (\ref{EtaShortTerm:Eq}) corresponds to an additive log-log
model for the cure probability, or equivalently, a complementary
log-log model for the long-term event probability.  

\section{Cure model with time-varying covariates} \label{TVcure:Sec}
\subsection{Model specification}
Assume now that the covariates are exogenous and can change values
over time.  Our proposal is to model the hazard rate at the population
level using
\begin{align}
  h_p(t|\mathbf{v}(t),\tilde{\mathbf{v}}(t)) &=
 \vartheta(\mathbf{v}(t)) f(t|\tilde{\mathbf{v}}(t))
  \nonumber \\
 &= \mathrm{e}^{\eta_\vartheta(\mathbf{v}(t))+\eta_F(\tilde{\mathbf{v}}(t))}
    f_0(t)S_0(t)^{\exp(\eta_F(\tilde{\mathbf{v}}(t))-1} ~,
  \label{TVmodel:PopHazard:Eq}
\end{align}
yielding a (promotion time) cure survival model with time-varying covariates,
shortly named the \ttf{TVcure} model.
The associated cumulative hazard function can be obtained numerically
using integration. In the special case where covariates are constant,
we recover the expressions for the population hazard and cumulative
hazard functions in \citet{Bremhorst:2016a}, see
(\ref{ExtendedPromotionTime:PopHazard:Eq}) and
(\ref{ExtendedPromotionTime:Pophazard:Eq}) in Section \ref{ExtendedPromotionTime:Sec},
with the parameter interpretation already discussed.
In the general case, the linear predictors $\eta_{\vartheta}$ and
$\eta_F$ not only change over units, but also potentially over
time. Therefore, the associated design matrices also depend on time
with
$\big(\eta_{\vartheta}(\mathbf{v}_{i}(t))\big)_{i=1}^n ={\mathbfcal
  X}_t \pmb{\psi}$, 
$\big(\eta_F(\tilde{\mathbf{v}}_{i}(t))\big)_{i=1}^n
={\tilde{\mathbfcal X}}_t \tilde{\pmb{\psi}}$.

Further assume that data for each unit can be reported in a regular manner
over time (measured in $dt$ units of time), such as with the monthly report
($dt=1$ month) of a woman status and her covariate values from age 20 ($t=0$) till
the event time (e.g. her first pregnancy) or the end of her follow-up
period (if she is childless by that time). Then, the data for the
$i$th unit would take the following form,
${\cal D}_i=\left\{(d_{it},\mathbf{v}_i(t),\tilde{\mathbf{v}}_i(t)):
  t=1,\ldots,t_i\right\}$, where $d_{it}$ is the event indicator
identically equal to 0 for all $t$, except perhaps the last value
$d_{it_i}$ equal to one if $\delta_i=1$ when an event is observed within
$(t_i-dt,t_i)$, and zero otherwise. 

\subsection{Inference} \label{Inference:Sec}
Assuming that the covariates remain
constant within a time unit $dt$, the conditional distribution of
$d_{it}$ for the $i$th subject still at risk at time $t-dt$
is approximately Poisson \citep{Lindsey1995} with mean
$\mu_{it}=h_p(t|\mathbf{v}(t),\tilde{\mathbf{v}}(t))\,dt$.
The log-likelihood contribution for that subject is
$$\ell_i(\pmb{\phi},\pmb{\psi},\tilde{\pmb{\psi}}|{\cal D}_i)
= -\sum_{t=1}^{t_i}\mu_{it}+d_{it_i}\log\mu_{it_i},$$
with the dependence of $\mu_{ij}$ on the three vectors of parameters
made explicit in Eqs.\,(\ref{f0:Bsplines:Eq}) to
(\ref{TVmodel:PopHazard:Eq}).
Smoothness priors for the spline parameters complete the model
description to counterbalance the flexibility
brought by the large B-spline bases in the specifications of $f_0(t)$
and of the additive terms in (\ref{EtaShortTerm:Eq})
and (\ref{EtaLongTerm:Eq}) \citep{FahrmeirLang2001},
\begin{align}
  \begin{split}
  p(\pmb{\phi}|\tau_0) &\propto \exp\left(-{1\over 2}\,
    \pmb{\phi}^\T (\tau_0\mathbf{P}_0)\, \pmb{\phi}\right)~, \\
  p(\pmb{\theta}_j|\lambda_j) &\propto \exp\left(-{1\over 2}\,
    \pmb{\theta}_j^\T (\lambda_j\mathbf{P})\, \pmb{\theta}_j \right)~~ (j=1,\ldots,J)~, \\
  p(\tilde{\pmb{\theta}}_j|\tilde\lambda_j) &\propto
    \exp\left(-{1\over 2}\,
    \tilde{\pmb{\theta}}_j^\T  (\tilde\lambda_j\tilde{\mathbf{P}})\, \tilde{\pmb{\theta}}_j \right)
  ~~(j=1,\ldots,\tilde{J}).
\end{split}
 \label{PriorTheta:Eq}
\end{align}
with Gamma priors for the penalty parameters,
$\tau_0,\lambda_j,\tilde{\lambda}_j\sim\G{1}{d=10^{-4}}$
\citep{LangBrezger2004}, or mixture of Gammas
\citep{JullionLambert:2007} with, as special cases,
half-Cauchy priors for the square-root of these parameters 
\citep{LambertBremhorst:2019}.
Assuming joint Normal priors for the 
parameters associated to the other covariates $\mathbf{z}$ and
$\tilde{\mathbf{z}}$,
$\pmb{\beta}
\sim \N{{\breve{b}}}{{\mathbf{Q}}^{-1}},~
\pmb{\gamma}
\sim \N{\breve{\mathbf{g}}}{{\tilde{\mathbf{Q}}}^{-1}},
$
the joint
priors for the regression and spline parameters in $\pmb{\psi}$ and
$\tilde{\pmb{\psi}}$ induce Gaussian Markov random fields (GMRF)
\citep{RueHeld2005} as they can be written as
\begin{align}
  \begin{split}
&
p(\pmb{\psi}|\pmb{\lambda}) \propto
\exp\left(-{1\over 2}~({\pmb{\psi}}-\mathbf{b})^\T \mathbf{K}_\lambda
({\pmb{\psi}}-\mathbf{b})
\right)~;~  \\
&
p(\tilde{\pmb{\psi}}|\tilde{\pmb{\lambda}}) \propto
\exp\left(-{1\over 2}~({\tilde{\pmb{\psi}}}-\mathbf{g})^\T \tilde{\mathbf{K}}_{\tilde{\lambda}}
({\tilde{\pmb{\psi}}}-\mathbf{g})
\right),~
\end{split}
                 \label{PriorPsi:Eq}
\end{align}
where~
$\mathbf{b}=(\breve{\mathbf{b}}^\T,\mathbf{0}_{JL}^\T)^\T$,
$\mathbf{K}_\lambda= \diag\big(\mathbf{Q},{\mathbfcal{P}}_\lambda\big)$,
${\mathbfcal{P}_\lambda}=\pmb{\Lambda} \kron \mathbf{P}$,
$[\pmb{\Lambda}]_{jj'}=\delta_{jj'}\lambda_j$,
$\mathbf{g}=(\breve{\mathbf{g}}^\T,\mathbf{0}_{\tilde{J}L}^\T)^\T$,
$\tilde{\mathbf{K}}_{\tilde{\lambda}}= \diag\big(\tilde{\mathbf{Q}},\tilde{{\mathbfcal{P}}}_{\tilde{\lambda}}\big)$,
${\tilde{\mathbfcal{P}}_\lambda}=\tilde{\pmb{\Lambda}} \kron \tilde{\mathbf{P}}$
and $[\tilde{\pmb{\Lambda}}]_{jj'}=\delta_{jj'}\tilde{\lambda}_j$.
Let $\pmb{\lambda}=(\lambda_j)^J_{j=1}$,
$\tilde{\pmb{\lambda}}=(\tilde{\lambda_j})^{\tilde{J}}_{j=1}$
with joint priors
$p(\pmb{\lambda})\propto \prod_jp(\lambda_j)$,
$p(\tilde{\pmb{\lambda}})\propto \prod_jp(\tilde{\lambda}_j)$.
If $\ell=\sum_{i=1}^n \ell_i$ denotes the log-likelihood, then
the joint posterior for the model parameters directly follows from
Bayes's theorem,
\begin{align*}
  &p(\pmb{\phi},\pmb{\psi},\tilde{\pmb{\psi}},
  \tau_0,{\pmb{\lambda}},\tilde{\pmb{\lambda}}|{\cal D})
    \propto    
   \\
  &\mbox{~~~~~}
    \exp\{\ell(\pmb{\phi},\pmb{\psi},\tilde{\pmb{\psi}}|{\cal D})\}
   \, p(\pmb{\phi}|\tau_0) p(\tau_0)
   \, p(\pmb{\psi}|\pmb{\lambda}) p(\pmb{\lambda}) 
   \, p(\tilde{\pmb{\psi}}|\tilde{\pmb{\lambda}}) p(\tilde{\pmb{\lambda}}).
\end{align*}

\subsubsection{Conditional estimation of the regression parameters}
The conditional posterior mode of the regression and spline parameters,
$\pmb{\zeta}=(\pmb{\psi}^\T,\tilde{\pmb{\psi}}^\T)^\T$ and $\pmb{\phi}$,
coincide with their
conditional penalized maximum likelihood estimates (PMLE) optimizing
\begin{align*}
  \ell_p
  &= \ell
    -{\tau_0\over 2}\,\pmb{\phi}^\T \mathbf{P}_0\, \pmb{\phi}
    -{1\over 2}~({\pmb{\psi}}-\mathbf{b})^\T \mathbf{K}_\lambda ({\pmb{\psi}}-\mathbf{b})    
    -{1\over 2}~({\tilde{\pmb{\psi}}}-\mathbf{g})^\T \tilde{\mathbf{K}}_{\tilde{\lambda}}
    ({\tilde{\pmb{\psi}}}-\mathbf{g}) \,.
\end{align*}
They can be obtained iteratively using the Newton-Raphson (N-R) algorithm)
built upon explicit forms for their respective gradient and precision
matrix, see 
Appendix \ref{Appendix:A}.
Practically, given values for $\pmb{\phi}$ and the penalty parameters
$\pmb{\tau}=(\pmb{\lambda}^\T,\tilde{\pmb{\lambda}}^\T)^\T$, 
repeat the following substitution till convergence:
\begin{align}
&\pmb{\zeta}
\longleftarrow
\pmb{\zeta}
- \mathcal{H}_\tau^{-1} \mathbf{U}_\tau^\zeta ~.
\label{zetaUpdate:Eq}
\end{align}
Estimates for the spline parameters $\pmb{\phi}$ defining $f_0(t)$ in
(\ref{f0:Bsplines:Eq}) are obtained in a similar manner, iteratively and conditionally on
the penalty parameter $\tau_0$,
\begin{align}
&{\pmb{\phi}}_{-k}
\longleftarrow
{\pmb{\phi}}_{-k}
- \left(({\pmb{H}}_{\tau_0}^{\phi\phi})_{-k,-k}\right)^{-1} ({\mathbf{U}}_{\tau_0}^\phi)_{-k} ~,
\label{phiUpdate:Eq}
\end{align}
with, for identification purposes, one of the vector components
arbitrarily set to zero, $\phi_{k}=0$.

\subsubsection{Selection of the penalty parameters} \label{SelectionPenaltyParams:Sec}
The marginal posterior for the penalty parameters
$\pmb{\tau}=(\pmb{\lambda}^\T,\tilde{\pmb{\lambda}}^\T)^\T$
tuning the
smoothness the additive terms can be obtained using the following
identity (with an implicit conditioning on $\pmb{\phi}$ and $\tau_0$),
\begin{align*}
  p(\pmb{\lambda},\tilde{\pmb{\lambda}}|{\cal D}) =
{ p(\pmb{\psi},\tilde{\pmb{\psi}},
  {\pmb{\lambda}},\tilde{\pmb{\lambda}}|{\cal D})
    / 
  p(\pmb{\psi},\tilde{\pmb{\psi}}|
  {\pmb{\lambda}},\tilde{\pmb{\lambda}},{\cal D})
  }~,
\end{align*}
with a Laplace's approximation substituted to the conditional posterior
of the spline parameters in the denominator, see
\citet{Lambert:2021} for a similar strategy in
nonparametric double additive location-scale models.
Evaluating that
expression at the conditional posterior modes yields the following approximation
to the marginal posterior of the penalty parameters,
\begin{align}
  p(\pmb{\lambda},\tilde{\pmb{\lambda}}|{\cal D})
  \stackrel{.}{\propto}
 p(\hat{\pmb{\psi}}_\lambda,\hat{\tilde{\pmb{\psi}}}_{\tilde{\lambda}},
  {\pmb{\lambda}},\tilde{\pmb{\lambda}}|{\cal D})
  \left|\Sigma_\tau^{-1}\right|^{-1/2}~,
  \label{MarginalPenalty:Eq}
\end{align}
where the blocks in the precision matrix
\begin{align} \label{SigmaGivenTau:Eq}
  \Sigma_\tau^{-1} = -\mathcal{H}_\tau =
\begin{bmatrix}
  -\mathbf{H}^{\psi\psi}_\lambda & -\mathbf{H}^{\psi\tilde{\psi}} \\
  -\mathbf{H}^{\tilde{\psi}\psi} & -\mathbf{H}^{\tilde{\psi}\tilde{\psi}}_{\tilde{\lambda}}
\end{bmatrix}  ~,
\end{align}
have explicit forms, see Appendix \ref{Appendix:A}. Maximizing that
marginal posterior enables to select the penalty parameters.
Remembering that the normalizing constants of the priors for
$\pmb{\psi}$ and $\tilde{\pmb{\psi}}$ in (\ref{PriorPsi:Eq}) depend on
the penalty parameters, the log of (\ref{MarginalPenalty:Eq}) can be
written (up to an additive constant) as
\begin{align}
  \begin{split}
 & \log p(\pmb{\lambda},\tilde{\pmb{\lambda}}|{\cal D})
  \stackrel{.}{=} \ell_p(\hat{\pmb{\phi}}_{\tau_0},\hat{\pmb{\zeta}}_{\tau}|\tau,{\cal D})
  + {1\over 2} \left\{
  \log |\mathbf{K}_\lambda|^+
  + \log |\tilde{\mathbf{K}}_{\tilde{\lambda}}|^+
  - \log |\mathcal{-H}_\tau|
  \right\}
  \\
 &~~~ = \ell_p(\hat{\pmb{\phi}}_{\tau_0},\hat{\pmb{\zeta}}_{\tau}|\tau,{\cal D})
 + {\rho(\mathbf{P})\over 2} \sum_{j=1}^J\log \lambda_j
 + {\rho(\tilde{\mathbf{P}})\over 2} \sum_{j=1}^{\tilde{J}}\log \tilde{\lambda}_j
    - {1\over 2}\log |\mathcal{-H}_\tau|~,
  \end{split}
    \label{logMarginalPostLambda:Eq}
\end{align}
where $|\mathbf{A}|^+$ denotes the product of the non-zero eigenvalues
of a semi-positive definite matrix $\mathbf{A}$.  Let
$\Sigma_\tau(\pmb{\theta}_{j})$ be the $L\times L$ submatrix in $\Sigma_\tau$
corresponding to the sub-vector $\pmb{\theta}_j$ in $(\pmb{\psi}^\T,\tilde{\pmb{\psi}}^\T)^\T$.
Given that
$$
{\partial \ell_p(\hat{\pmb{\phi}}_{\tau_0},\hat{\pmb{\zeta}}_{\tau}|\tau,{\cal D})
  \over \partial \lambda_j}
=
{\partial \ell_p(\hat{\pmb{\phi}}_{\tau_0},\hat{\pmb{\zeta}}_{\tau}|\tau,{\cal D})
  \over \partial \hat{\pmb{\psi}}_\lambda^\T} {\partial \hat{\pmb{\psi}}_\lambda \over \partial \lambda_j}
-{1\over 2}\,
\hat{\pmb{\theta}}_{j\lambda}^\T \mathbf{P} \hat{\pmb{\theta}}_{j\lambda}
~~~~(1\leq j\leq J)
$$
with the first factor in that expression equal to zero, and remembering that
${\partial \log|A_s| / \partial s}=\tr{A_s^{-1} {\partial A_s / \partial s}}$
for a positive definite matrix $A_s$,
one has
\begin{align}
  \begin{split}
    {\partial \log p(\pmb{\lambda},\tilde{\pmb{\lambda}}|{\cal D}) \over \partial \lambda_j}
    &= {1\over 2} \left\{
      {\rho(\mathbf{P})\over \lambda_j}
      -\hat{\pmb{\theta}}_{j\lambda}^\T \mathbf{P} \hat{\pmb{\theta}}_{j\lambda}
      - \tr{\Sigma_\tau(\pmb{\theta}_j)\mathbf{P}}
      \right\}.
    \end{split}
      \label{Dlogplambda:Eq}
\end{align}
A similar expression (with $1\leq j\leq \tilde{J}$) can be obtained for
\begin{align}
  \begin{split}
    {\partial \log p(\pmb{\lambda},\tilde{\pmb{\lambda}}|{\cal D}) \over \partial \tilde{\lambda}_j}
    &= {1\over 2} \left\{
      {\rho(\tilde{\mathbf{P}})\over \tilde{\lambda}_j}
      -\hat{\tilde{\pmb{\theta}}}_{j\tilde{\lambda}}^\T \tilde{\mathbf{P}} \hat{\tilde{\pmb{\theta}}}_{j\tilde{\lambda}}
      - \tr{\Sigma_\tau(\tilde{\pmb{\theta}}_j)\tilde{\mathbf{P}}}
      \right\}.
    \end{split}
      \label{Dlogplambdatilde:Eq}
\end{align}
The MAP estimate for $\pmb{\lambda}$ and $\tilde{\pmb{\lambda}}$ are the solutions of
(\ref{Dlogplambda:Eq}) and (\ref{Dlogplambdatilde:Eq}) set to zero for all $j$.
This can be done using the fixed point method with the following substitutions
iterated till convergence:
\begin{align}
  \begin{split}
    &\lambda_j^{-1} \longleftarrow
  {\hat{\pmb{\theta}}_{j\lambda}^\T \mathbf{P} \hat{\pmb{\theta}}_{j\lambda}
    + \tr{\Sigma_\tau(\pmb{\theta}_j)\mathbf{P}} \over \rho(\mathbf{P})}
  = {\E(\pmb{\theta}_j^\T\mathbf{P}\pmb{\theta}_j|\pmb{\lambda},{\cal D}) 
     \over \rho(\mathbf{P})} 
  ~~~~~(1\leq j\leq J) \\
 & \tilde{\lambda}_j^{-1} \longleftarrow
  {\hat{\tilde{\pmb{\theta}}}_{j\tilde{\lambda}}^\T \tilde{\mathbf{P}} \hat{\tilde{\pmb{\theta}}}_{j\tilde{\lambda}}
  + \tr{\Sigma_\tau(\tilde{\pmb{\theta}}_j)\tilde{\mathbf{P}}} \over \rho(\tilde{\mathbf{P}})}
  =  {\E(\tilde{\pmb{\theta}}_j^\T\tilde{\mathbf{P}}\tilde{\pmb{\theta}}_j|\tilde{\pmb{\lambda}},{\cal D}) \over \rho(\tilde{\mathbf{P}})}
~~~~(1\leq j\leq \tilde{J})  \,. 
\end{split}
   \label{LambdaUpdate:Eq}
\end{align}
The connection to conditional expectations in (\ref{LambdaUpdate:Eq})
results from the preceding Laplace approximations to
$(\pmb{\theta}|\pmb{\lambda},{\cal D})$ and
$(\tilde{\pmb{\theta}}|\tilde{\pmb{\lambda}},{\cal D})$.  The combination of
(\ref{zetaUpdate:Eq}) and (\ref{LambdaUpdate:Eq}) leads to Algorithm
\ref{Algorithm1} for the selection of $\pmb{\tau}$ and the estimation
of $\pmb{\zeta}$ (for a given value of $\pmb{\phi}$).
\begin{algorithm} 
    \caption{Selection of $\pmb{\tau}$ and estimation of ${\pmb{\zeta}}$ (for given $\pmb{\phi}$)}
  \label{Algorithm1}
  \SetKwInput{Input}{Input}\SetKwInput{Output}{Output}

  \Input{Spline parameters $\pmb{\phi}$ from the short-term survival submodel,
    data ${\cal D}=\cup_{i=1}^n\left\{(d_{it},\mathbf{v}_i(t),\tilde{\mathbf{v}}_i(t)):
      t=1,\ldots,t_i\right\}$.}    
  \Output{Selected $\pmb{\tau}=(\pmb{\lambda}^\T,\tilde{\pmb{\lambda}}^\T)^\T$ and
    estimated ${\hat{\pmb{\zeta}}}_\tau$
    (given $\pmb{\phi}$).}
  {\bf Note:} $\pmb{\zeta}=({\pmb{\psi}}^\T,{\tilde{\pmb{\psi}}}^\T)^\T$ with
  $\pmb{\psi} = \begin{pmatrix}\pmb{\beta}, \Vec{\mathbf{\Theta}}\end{pmatrix}$,
  $\tilde{\pmb{\psi}} = \begin{pmatrix}\pmb{\gamma}^\T, \Vec{\tilde{\mathbf{\Theta}}}^\T\end{pmatrix}^\T$
  \BlankLine
  
  \Repeat{convergence}{
    \Repeat{$||\mathbf{U}_\tau^\zeta||_\infty < \epsilon$\, }{
      Evaluate $\mathbf{U}_\tau^\zeta$ and $\mathcal{H}_\tau$
      using (\ref{GradHessPsi:Eq}), (\ref{GradHesPsiTilde:Eq}) \& (\ref{GradHessZeta:Eq}). \\
      Update ~$\pmb{\zeta} \longleftarrow \pmb{\zeta} - \mathcal{H}_\tau^{-1} \mathbf{U}_\tau^\zeta$
    }
    \Repeat{convergence }{
      $\lambda_j^{-1} \longleftarrow
      \left\{{{\pmb{\theta}}_{j}^\T \mathbf{P} {\pmb{\theta}}_{j}
          + \tr{\Sigma_\tau(\pmb{\theta}_j)\mathbf{P}}} \right\} / \rho(\mathbf{P})$ ~~~($1\leq j\leq J$) \\
      $\tilde{\lambda}_j^{-1} \longleftarrow
      \left\{{\tilde{\pmb{\theta}}}_{j}^\T \tilde{\mathbf{P}} {\tilde{\pmb{\theta}}}_{j}
        + \tr{\Sigma_\tau(\tilde{\pmb{\theta}}_j)\tilde{\mathbf{P}}} \right\} / \rho(\tilde{\mathbf{P}})$
      ~~($1\leq j\leq \tilde{J}$) \\
      Update $\Sigma_\tau=(-\mathcal{H}_\tau)^{-1}$ using (\ref{GradHessPsi:Eq}), (\ref{GradHesPsiTilde:Eq}) \& (\ref{GradHessZeta:Eq}).
  }
}
\end{algorithm}

The selection of $\tau_0$ tuning the smoothness of $f_0(t)$ proceeds
in a similar way. Conditionally on the regression parameters, the
maximization of $p(\tau_0|{\cal D})$ (approximated using the same type of
arguments as with $p(\pmb{\lambda},\tilde{\pmb{\lambda}}|{\cal D})$)
can be made using the following substitution repeated till
convergence:
\begin{align}
  &\tau_0^{-1} \longleftarrow
  {\hat{\pmb{\phi}}_{\tau_0}^\T \mathbf{P}_0 \hat{\pmb{\phi}}_{\tau_0}
    + \tr{(-\mathbf{H}_{\tau_0}^{\phi\phi})^{-1}\mathbf{P}_0} \over \rho(\mathbf{P}_0)}
  =  {\E(\pmb{\phi}^\T\mathbf{P}_0\pmb{\phi}|\tau_0,{\cal D}) 
         \over \rho(\mathbf{P}_0)}
    \label{OmegaUpdate:Eq}
\end{align}
The combination of (\ref{phiUpdate:Eq}) and (\ref{OmegaUpdate:Eq}) leads to Algorithm
\ref{Algorithm2} for the selection of $\tau_0$ and the estimation
of $\pmb{\phi}$ (for a given value of $\pmb{\zeta}$).

\begin{algorithm} 
  \SetKwInput{Input}{Input}\SetKwInput{Output}{Output}

  \Input{Regression and spline parameters $\pmb{\zeta}$, 
    data~${\cal D}=\cup_{i=1}^n\left\{(d_{it},\mathbf{v}_i(t),\tilde{\mathbf{v}}_i(t)):
      t=1,\ldots,t_i\right\}$.}
  \Output{Selected $\tau_0$ and estimated ${\hat{\pmb{\phi}}}_{\tau_0}$
     (given $\pmb{\zeta}$).}
  \BlankLine
  
  \Repeat{convergence}{
    \Repeat{$||\mathbf{U}_{\tau_0}^\phi||_\infty < \epsilon$\, }{
      Evaluate $\mathbf{U}_{\tau_0}^\phi$ and $\mathbf{H}_{\tau_0}^{\phi\phi}$ using (\ref{GradHesPhi:Eq}) \\
      Update ${\pmb{\phi}}_{-k} \longleftarrow
      {\pmb{\phi}}_{-k} - \left(({\pmb{H}}_{\tau_0}^{\phi\phi})_{-k,-k}\right)^{-1} ({\mathbf{U}}_{\tau_0}^\phi)_{-k}$
      with $\phi_{k}=0$.
    }
    \Repeat{convergence }{
      Update \,$\tau_0^{-1} \longleftarrow
  \left\{{\pmb{\phi}}^\T \mathbf{P}_0 {\pmb{\phi}}
    + \tr{(-\mathbf{H}_{\tau_0}^{\phi\phi})^{-1}\mathbf{P}_0}\right\} / \rho(\mathbf{P}_0)$ \\
      Update $\mathbf{H}_{\tau_0}^{\phi\phi}$ using (\ref{GradHesPhi:Eq}).
  }
}
  \caption{Selection of $\tau_0$ and estimation of ${\pmb{\phi}}$ (for given $\pmb{\zeta}$)}
  \label{Algorithm2}
\end{algorithm}

\subsubsection{Global estimation algorithm}
The double additive TVcure model with time-varying covariates can be
fitted using Algorithm \ref{Algorithm3}. It alternates (till
convergence) the selection and estimation of parameters
$\{\pmb{\tau}, \pmb{\zeta}\}$ in the regression submodels, with that of
the parameters $\{\tau_0,\pmb{\phi}\}$ specifying the baseline
short-term survival dynamics (for reference values of the categorical
covariates and additive terms set to zero). Possible initial values to
initiate the algorithm are 0 for all the spline and regression
parameters in $\pmb{\zeta}$ and $\pmb{\phi}$, and moderately large
values (100, say) for all the penalty parameters in $\pmb{\tau}$ and
$\tau_0$.  The procedure was implemented using pure R code in a
package named \ttf{tvcure} maintained by the author.  Convergence is
fast and only takes a couple of seconds using a basic laptop computer.

\begin{algorithm} 
  {
  \SetKwInput{Input}{Input}\SetKwInput{Output}{Output}
  \SetKwInput{Goal}{Goal}

  \Goal{Fit of the TVcure model described in Section \ref{TVcure:Sec}.}
  \Input{Data~${\cal D}=\cup_{i=1}^n\left\{(d_{it},\mathbf{v}_i(t),\tilde{\mathbf{v}}_i(t)):
      t=1,\ldots,t_i\right\}$ and a TVcure model specification.}
  \Output{Estimates for $\pmb{\zeta}=({\pmb{\psi}}^\T,{\tilde{\pmb{\psi}}}^\T)^\T$ and $\pmb{\phi}$
    for the selected penalty parameters $\pmb{\tau}=({\pmb{\lambda}}^\T, \tilde{\pmb{\lambda}}^\T)^\T$ and $\tau_0$.}
  \BlankLine

  \Repeat{convergence}{
    - Select $\tau_0$ and estimate ${\pmb{\phi}}$ (given
    $\pmb{\zeta}$ and ${\cal D}$) using Algorithm \ref{Algorithm2}.\\
    - Select $\pmb{\tau}$ and estimate ${\pmb{\zeta}}$ (given $\pmb{\phi}$ and
    ${\cal D}$) using Algorithm \ref{Algorithm1}.
  }
  \caption{Fitting the double additive cure model with time-varying covariates (TVcure)}
  \label{Algorithm3}
  }
\end{algorithm}

\section{Simulation study} \label{SimulationStudy:Sec}

A simulation study was setup to evaluate the ability of the algorithms
described in Section \ref{TVcure:Sec} to estimate the different
ingredients of the TVcure model from right-censored data, including
the additive terms and the reference cumulative hazard function
$\mathrm{e}^{\beta_0}F_0(t)$. With the application from Section
\ref{Application:Sec} in mind, 
$S=500$ datasets of size $n=500$ or $1500$ were simulated using the
data generating mechanism corresponding to the extended promotion time
model with population hazard function
(\ref{ExtendedPromotionTime:Pophazard:Eq}), where $F_0(t)=1-S_0(t)$
with $S_0(t)$ given by
the survival function of a Weibull with shape parameter $2.65$ and
scale parameter $133$. The regression parameters in
(\ref{EtaShortTerm:Eq}) and (\ref{EtaLongTerm:Eq}) were taken to be
$\beta_0=0$, $\beta_1=-.1$, $\beta_2=.15$ and $\gamma_1=.1$,
$\gamma_2=.2$,
with independent Bernoulli or Normally distributed covariates, 
$z_1,z_3 \sim .5\,\text{Bern}(.5)$, $z_2,z_4 \sim\N{0}{1}$.  The
following additive terms were considered,
\begin{align*}
    f_1(x_1)&= -1.14+2.4 x_1 -.88 x_1^2 ~;~
    f_2(x_2)= -.3 \cos(2\pi x_2) \\
    \tilde{f}_1(x_1)&= .15 - .5 \cos(\pi(x_1-.75)) ~;~
    \tilde{f}_2(x_3)= .6(x_3-.5)
\end{align*}
with $x_1\sim \text{Uniform}{(0,1.5)}$ shared by the long- and
short-term survival submodels, while
$x_2, x_3\sim \text{Uniform}{(0,1)}$ are independent covariates
specific to each of the preceding submodels. Such data can be
generated using the biological motivation of the promotion time model
given by \citet{Yakovlev:1996} with
$(n_i|\mathbf{v}_i)\sim\text{Pois}(\vartheta(\mathbf{v}_i))$,
$y_i=+\infty$ if $n_i=0$ (in which case unit $i$ is `cured') and
$y_i=\min\{\breve{y}_{im}:m=1,\ldots,n_i\}$ otherwise, where
$(\breve{y}_{im}|\tilde{\mathbf{v}})$ are independently and
identically distributed random variables with
c.d.f. $F(\cdot|\tilde{\mathbf{v}})$.  Independent censoring times
were generated using a Uniform on $(120,299)$ (Scenario 1) or
$(60,299)$ (Scenario 2), yielding
an observed (large) right-censoring rate close to 44\% or 51\%, including a
marginal (unknown) cure rate of 40\%. If $y_i$ denotes the generated
non-censored response for unit $i$ ($i=1,\ldots,n$), then the observed
response is $(t_i,\delta_i)$ where $t_i=\min\{y_i,c_i\}$ and
$\delta_i=I(y_i<c_i)$ is the event indicator. These data are
transformed in a person-month format in a second step, see Section
\ref{Inference:Sec}, with $dt=1$ (month) and event indicator $d_{it}$
for a unit $i$ still at risk after a follow-up of $t$ months ($t=1,\ldots,299$).

\begin{figure}[!ht]\centering
\includegraphics[width=\textwidth]{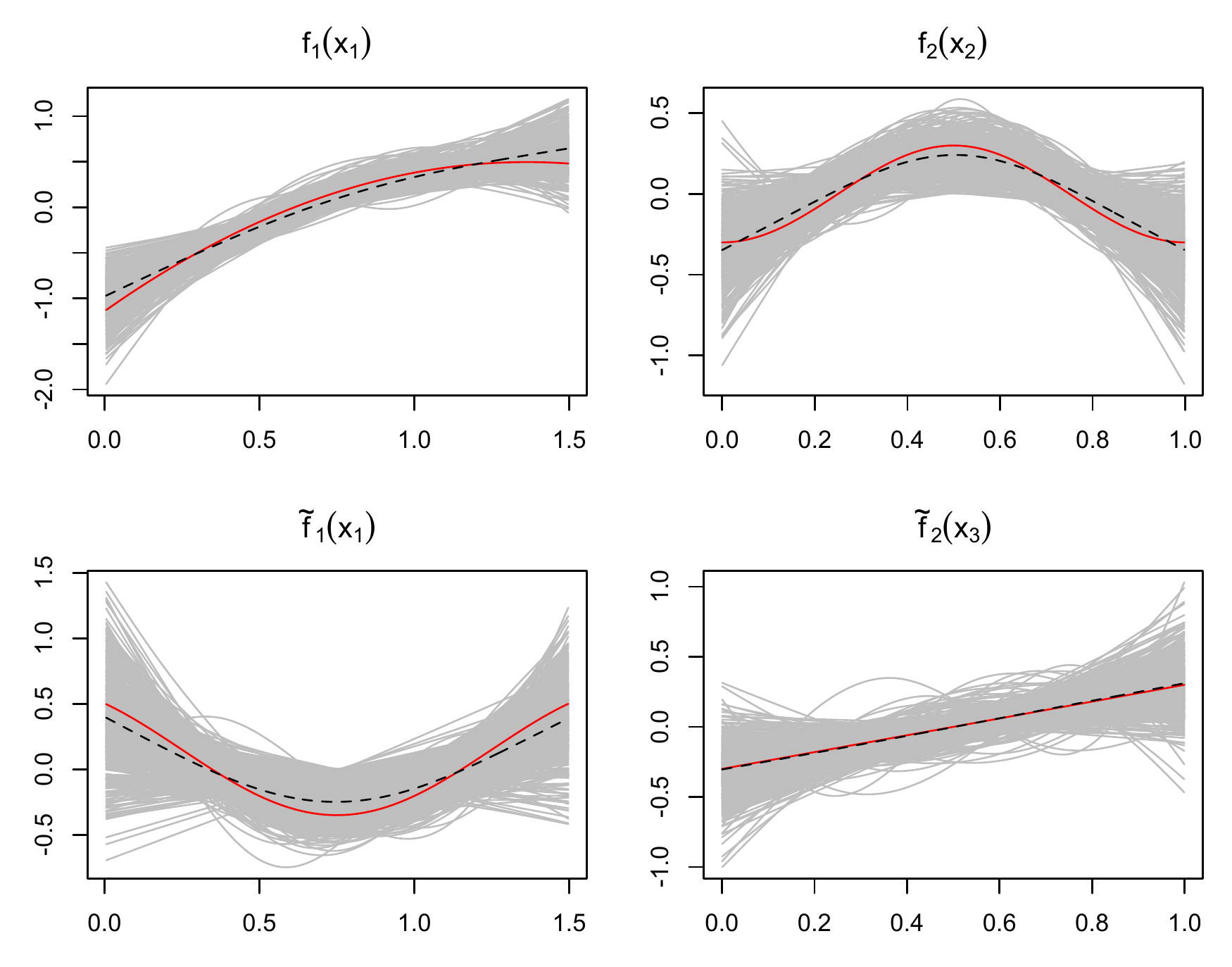}
\caption{\label{simulationAdditiveTermsN500RC44:Fig} Simulation study
  ($n=500$ - Scenario 2):
  Estimated additive terms (in grey) for each of the $S=500$ datasets
  and their average values (dashed curves) over the $S$ replicates with
  the solid curves (in red) corresponding to the `true' additive
  functions.} 
\end{figure}

\begin{figure}[!ht]\centering
\includegraphics[width=.49\textwidth]{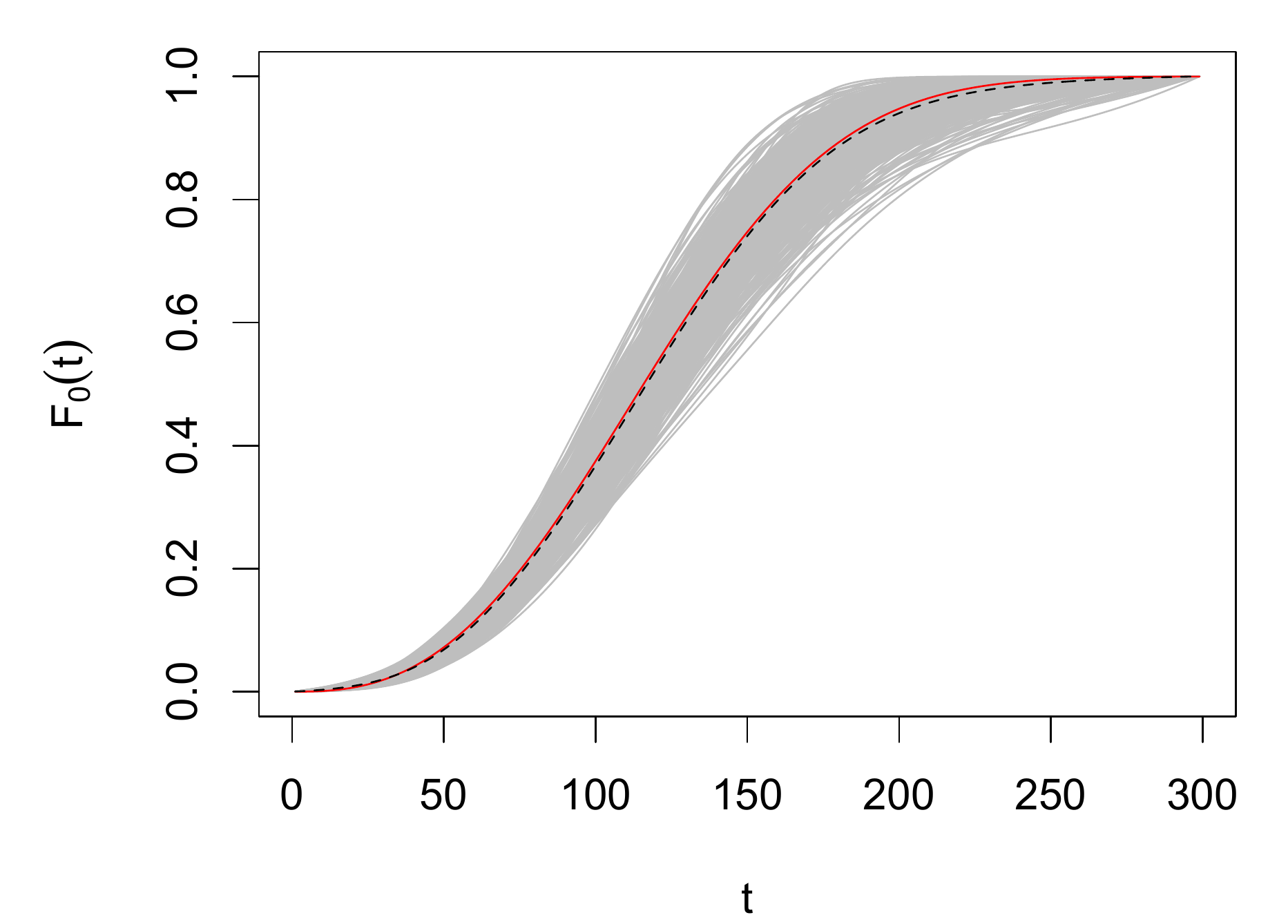}
\includegraphics[width=.49\textwidth]{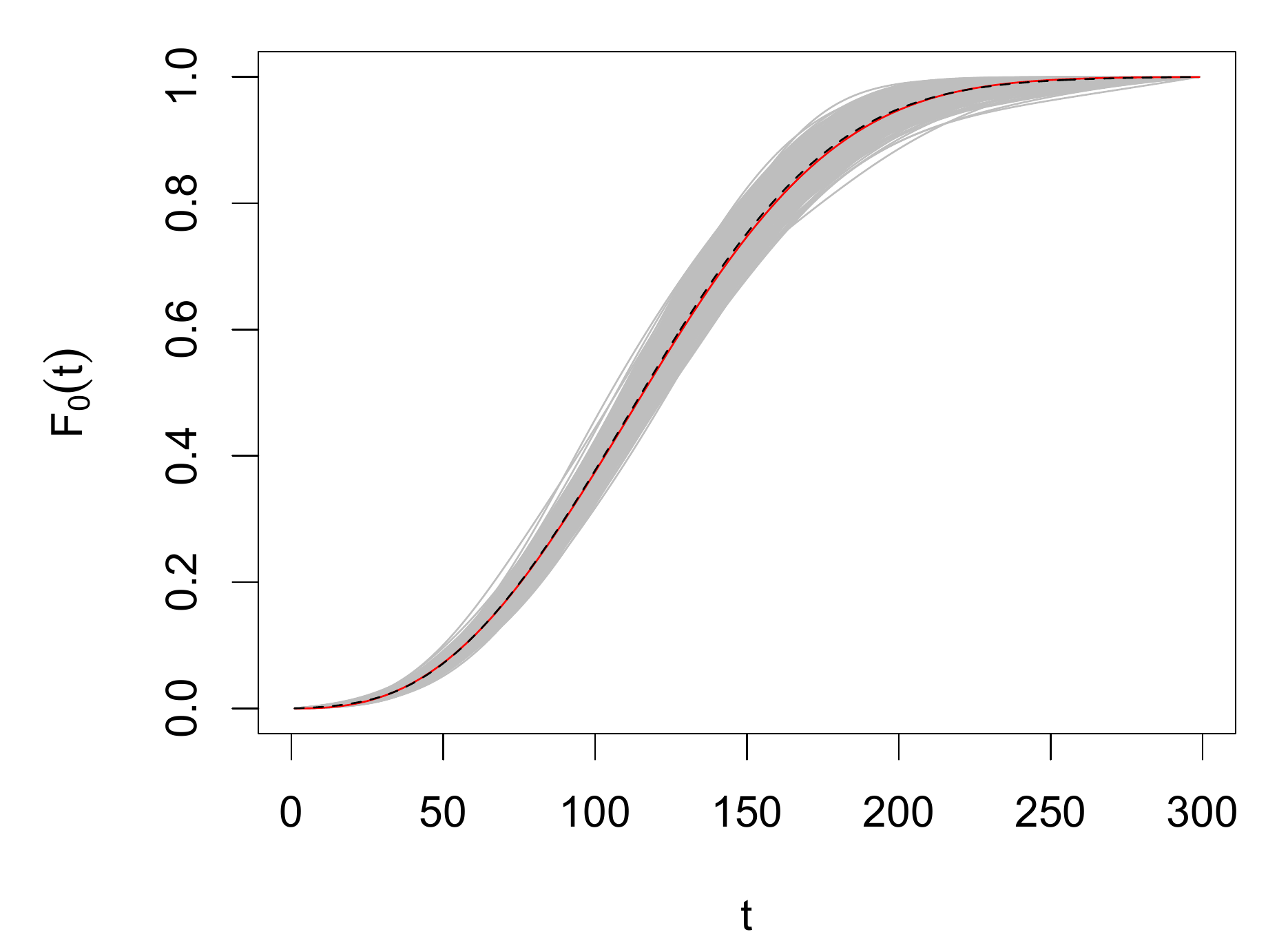}
\caption{\label{simulationF0RC44:Fig}Simulation study (Scenario 2):
  Estimated cumulative baseline hazard function $F_0(t)$ for each of the $S=500$ datasets
  and their average values (dashed curves) over the $S$ replicates with
  the solid curves (in red) corresponding to the `true' $F_0(t)$
  (Left panel: $n=500$ ; Right panel: $n=1500$).}
\end{figure}

The estimated additive terms for the $S=500$ datasets and their
average values over these $S$ replicates can be found in
Fig.\,\ref{simulationAdditiveTermsN500RC44:Fig} for the least
favorable setting (Scenario 2 when $n=500$) and with
$K=10$ (penalized) B-splines associated to equidistant knots spanning
the observed range for the associated covariate. The shape of the
additive terms is quite well estimated despite the moderate sample
size, the large proportion of right-censored data
and the sophistication of the model. Summary information on
the quality of the estimation of the additive terms are reported in
Table \ref{simulationAdditiveTermsRC44:Tab}.
Not surprisingly, the reconstruction improves with the sample
size and with decreasing right censoring rates.
The mean absolute bias
and the root mean integrated squared error (RMISE) decrease with
sample size, proportionally to $n^{-1/2}$ for the RMISE. The effective
(mean) coverages of pointwise 95\% credible intervals for the additive
terms are close to their nominal value. Similar conclusions can be
reached for the estimation of the regression parameters, see Table
\ref{simulationRegressionParmsRC44:Tab}.  Finally, the estimates of
the standardized cumulative hazard function $F_0(t)$ for the $S$
simulated datasets of size $n=500$ can be found in
Fig.~\ref{simulationF0RC44:Fig} with also very satisfactory
results.

\begin{table}
  \caption{\label{simulationAdditiveTermsRC44:Tab} Simulation study
    (Scenario 2): estimation of the additive terms in the long- and
    short-term submodels. Mean absolute bias, Root mean integrated
    squared error (RMISE) and mean effective coverage of 95\%
    pointwise credible intervals.} \mbox{}  
  \centering
    \begin{tabular}{lclrrrr}
      \toprule
Censoring &$n$&& $f_1(x)$ & $f_2(x)$ & $\tilde{f}_1(x)$ & $\tilde{f}_2(x)$ \\
  \hline
Scenario 1  &  $500$ & MA-Bias & 0.097 & 0.087 & 0.131 & 0.068 \\
          && RMISE   & 0.127 & 0.112 & 0.174 & 0.097 \\
          && Coverage & 81.2 & 90.7 & 87.3 & 96.0 \\
& $1500$ & MA-Bias & 0.053 & 0.053 & 0.070 & 0.042 \\
          && RMISE   & 0.071 & 0.069 & 0.092 & 0.057 \\
          && Coverage & 93.1 & 93.3 & 94.5 & 96.0\\
Scenario 2 &  $500$ & MA-Bias & 0.107 & 0.091 & 0.110 & 0.073 \\
          && RMISE   & 0.139 & 0.118 & 0.186 & 0.105 \\
          && Coverage & 80.2 & 90.5 & 86.6 & 96.2 \\
& $1500$ & MA-Bias & 0.057 & 0.055 & 0.075 & 0.040 \\
          && RMISE   & 0.077 & 0.072 & 0.010 & 0.056 \\
          && Coverage & 93.0 & 93.4 & 94.8 & 96.7\\
      \bottomrule
    \end{tabular}
  \end{table}

\begin{table}
\caption{\label{simulationRegressionParmsRC44:Tab} Simulation study
  (Scenario 2): 
  estimation of regression parameters in the long- and short-term
  submodels.
  Bias, RMSE and effective coverage of 95\% credible intervals.} \mbox{}
  \centering
    \begin{tabular}{lclrrrrrrr}
      \toprule
  &&& $\beta_0$ & $\beta_1$ & $\beta_2$ & $\gamma_1$ & $\gamma_2$ \\
Censoring &$n$&  & 0.000 & -0.100 & 0.150 & 0.100 & 0.200 \\
  \hline
Scenario 1 & $500$ &Bias  & 0.007 & -0.006 & 0.010 & 0.007 & 0.005 \\
         &&RMSE  & 0.102 &  0.065 & 0.064 & 0.083 & 0.078 \\
         && Coverage & 89.4 & 92.8   & 93.8  & 81.6  & 95.8\\
         &$1500$&Bias  &-0.008 &  0.000 & 0.001 & 0.000 & 0.003 \\
         &&RMSE  & 0.053 &  0.035 & 0.037 & 0.045 & 0.045 \\
         &&Coverage & 93.2 & 94.4 & 93.8  & 85.4  & 96.0   \\
Scenario 2 & $500$ &Bias  & 0.007 &-0.008 & 0.006 & 0.016 & 0.002 \\
         &&RMSE  & 0.110 &  0.067 & 0.068 & 0.089 & 0.085 \\
         && Coverage & 90.4 & 94.4& 94.2  & 83.8  & 95.4\\
         &$1500$&Bias  &-0.011 &  0.002 & 0.000 & 0.002 & 0.005 \\
         &&RMSE  & 0.061 &  0.038 & 0.037 & 0.046 & 0.051 \\
         &&Coverage & 91.6 & 94.4 & 95.4  & 85.4  & 93.8 \\
\bottomrule
    \end{tabular}
\end{table}

\section{Application: women's earnings and fertility in Germany} \label{Application:Sec}

In this section we illustrate how the method works with real data.
For this purpose, we draw on register data from Germany and study how
women's earnings relate to first birth behavior. This topic is an
ideal test case to illustrate the method that was developed in this
paper. First, Germany is a country with one of the highest shares of
ever childless women in Europe \citep{KreyenfeldKonietzka2017}. Thus,
there is always a high `cure fraction' in each cohort. Second, Germany
has enacted several major family policy reforms in recent years. A
very important reform has been the parental leave benefit reform,
enacted in 2007. Compared to prior regulations, it sets stronger
incentives than before to establish in the labour market before having
children.  Thus, one might expect the association between women's
earnings and fertility to be stronger for younger than for older cohorts.

The data of interest is a random sample from the German Pension
registers of years 2017 and 2019 (Data extract SUFVSKT 2017/SUFVSKT
2019). The German pension registers cover roughly 90\% of the resident
population in Germany. Certain professions (farmers, lawyers) and
civil servants are not included in the data.
For the subsequent analysis, we have limited the investigation to West
German women of the birth cohorts 1950-74. East Germany is eliminated
from this study, as fertility patterns in the two parts of the country
were rather different, in particular before reunification. We focus
furthermore on women who did not have any children yet at age
20. Thus, teenage fertility is not part of this investigation. The
sample size comprises 15,248 women and 11,019 first births.

The outcome of interest, the starting month of the pregnancy, was
calculated as the birth date of the first child minus 9 months. The
follow-up considered for each woman started at the age of 20 until at
most 45 with a possible interruption at the first pregnancy or due to
a loss of follow-up (i.e.~right-censoring) with, in the latter case,
an uncertainty on the `cure' (i.e.~childless) final status of the
person. Thus, in this fertility context, a woman will be considered
`cured' if she doesn't have a child by age 45.
The dataset also contains complete monthly employment and earning
histories of women. Earnings are stored as earning points, where one
earning point represents the average annual earnings in a given
year. Note that we only observe earnings during regular
employment. Regular employment is employment that results into pension
credits to the German pension fund. Women who do not receive any
earnings from regular employment enter the analysis with `zero'
earnings. These women may be studying, unemployed or out of the labour
market for other reasons. For the main parts of the investigation, we
use earnings as a continuous covariate. A set of time-varying
`employment variables' control for whether a woman is studying,
employed, registered unemployed or not in regular employment for other
reasons.  We also control for birth cohort where we distinguish
between cohorts born 1950-54, 1955-59, 1960-64, 1965-69,
1970-74. Table \ref{App:SummaryStats:Tab} includes descriptive
statistics by groups of birth cohorts.
The women described here as `right-censored' had their follow-up
interrupted before their first potential pregnancy by age 45: thus,
there is uncertainty about their maternal status at age 45. The table
suggests that the proportion of West German women childless by age 45
is growing, starting from about 21\% for the 1950-55 cohort to 28\%
for the 1970-74 cohort. Uncertainty remains, however, given the
decreasing percentage of complete follow-ups.
\begin{table}
  \caption{\label{App:SummaryStats:Tab} Summary statistics on the
    follow-up of West German women's cohorts and the number of months
    between their 20th birthday and their first pregnancy (Source:
    SUFVSKT 2018).} 
  \centering
\begin{tabular}{lccrrr}
\toprule
  &     &  Person             & \multicolumn{3}{c}{Mother by Age 45} \\
  \cmidrule(lr){4-6}
Cohort & $n$ & months & \multicolumn{1}{c}{Yes} &  \multicolumn{1}{c}{No} & Right-cens. \\
  \midrule
  1950-54    & 2423 & 277668 & 1900 (78.4\%)  & 519 (21.4\%) & ~\,4 (0.2\%)  \\
  1955-59    & 2388 & 313692 & 1762 (73.8\%)  & 613 (25.7\%) & 13 (0.5\%)  \\
  1960-64    & 3029 & 431144 & 2186 (72.2\%)  & 797 (26.3\%) & 46 (1.5\%)  \\
  1965-69    & 3385 & 511119 & 2410 (71.2\%)  & 887 (26.2\%) & 88 (2.6\%)  \\  
  1970-74    & 4023 & 655489 & 2761 (68.6\%)  & 1130 (28.1\%) & 132 (3.3\%)  \\  
\bottomrule[0.09 em]
   \end{tabular}
\end{table}

\begin{figure}[!ht]\centering
\includegraphics[width=.6\textwidth]{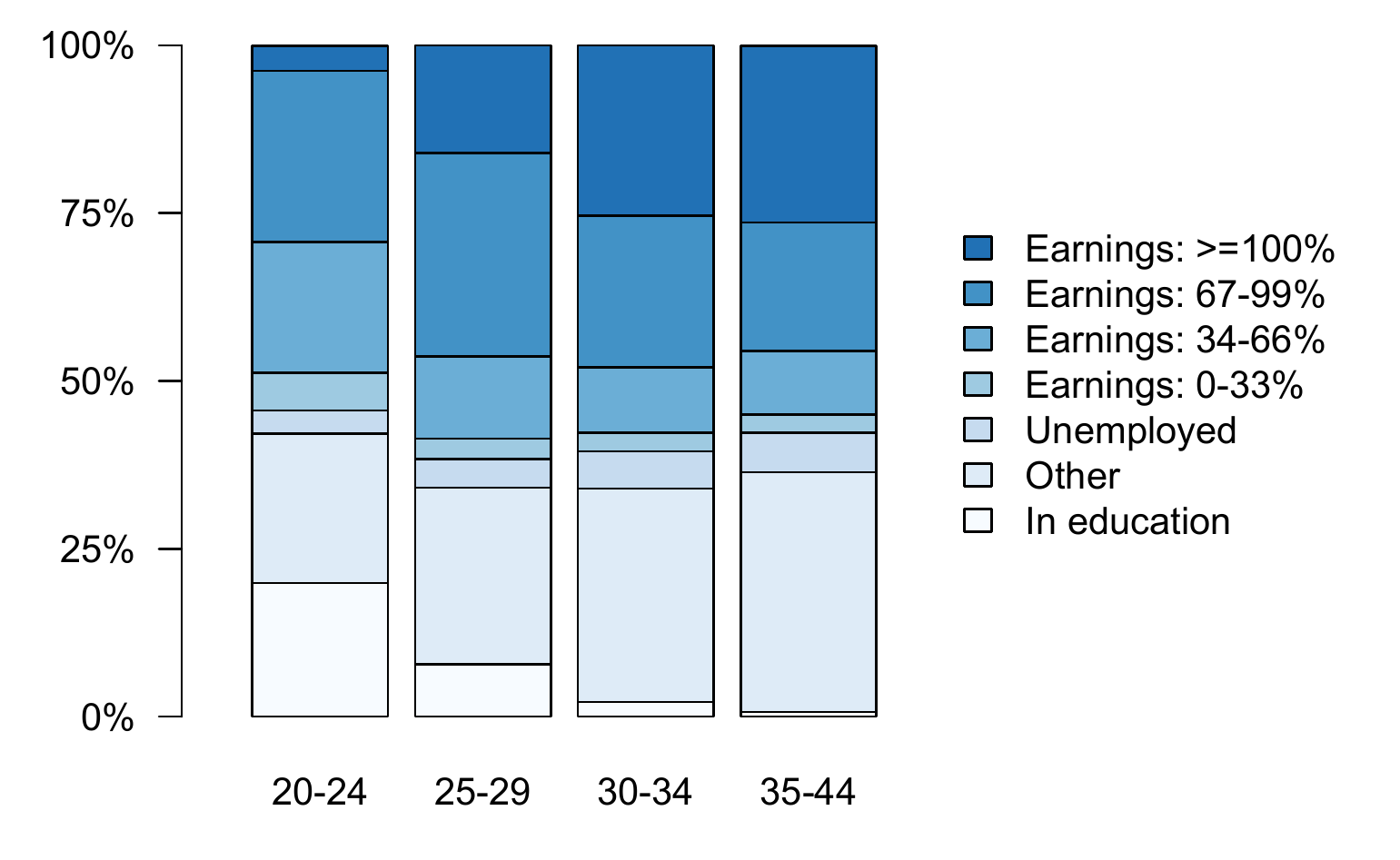}
\caption{\label{EarningsByAge:Fig}Earnings and employment status of
  nulliparous women, cohorts 1950-74, West Germany.}
\end{figure}

For robustness checks, we also conducted an additional analysis
where we grouped earnings into four categories $(0,.33]$ (`Low'),
$(.33,.66]$ (`Medium'), $(.66,1.0)$ (`High') and $(1.00,1.5]$ (`Top').
As earnings are measured in earning points, the cut-point represent
women who earn less than 33\% of average earnings, those between
33-66\%, those between 66\% and average earnings. The last category
comprises women who earn more than average earnings. Figure
\ref{EarningsByAge:Fig} plots the distribution of the categorized
outcome variable. As expected, a large fraction of the younger women
(ages 20--24) are in education. Noteworthy is the relatively high
share of women who do not receive any earnings from regular
employment. Further, only a small share of 25\% earns more than
average at age $30$ and older.
\begin{figure}[!ht]\centering
\includegraphics[width=.49\textwidth]{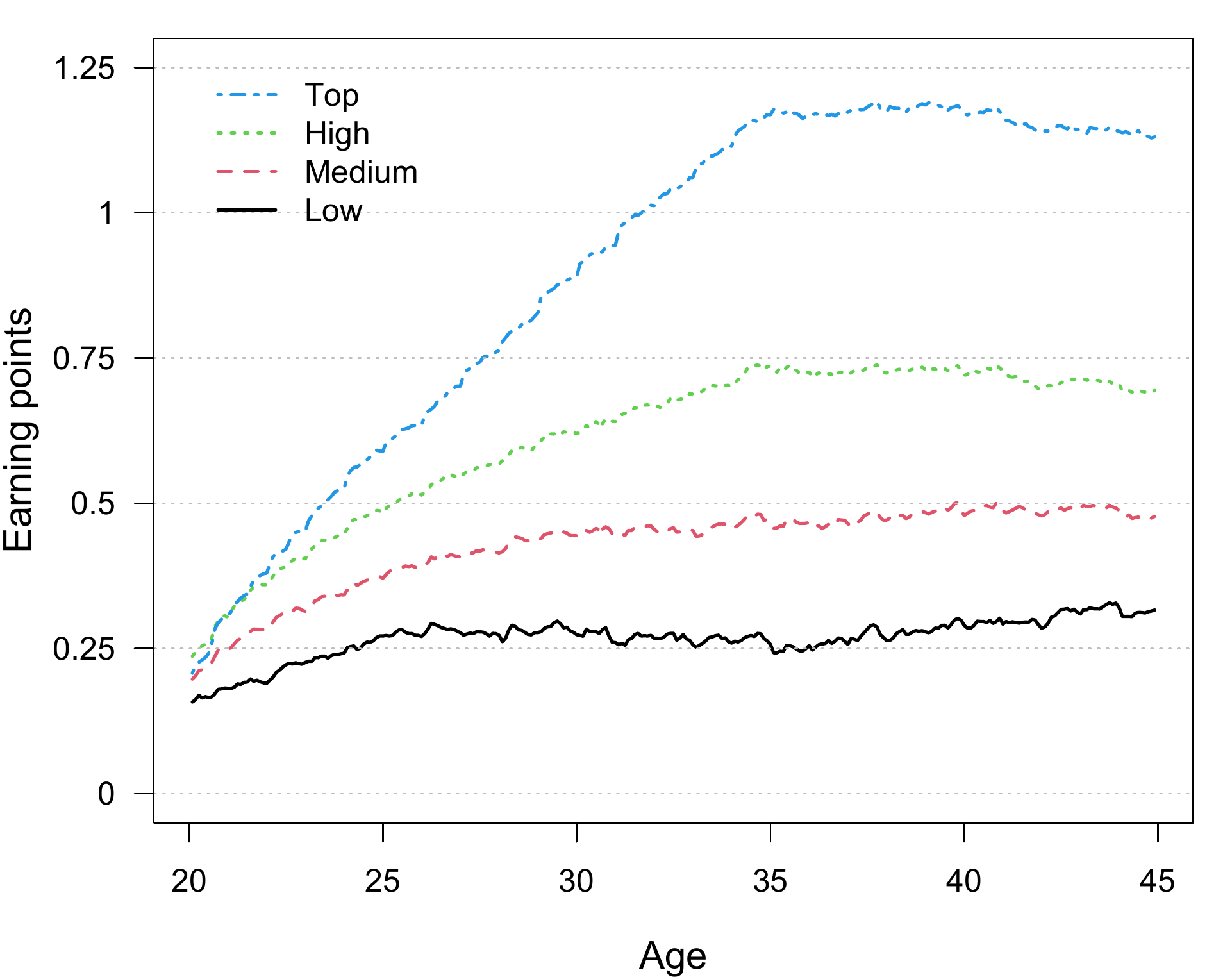}
\includegraphics[width=.49\textwidth]{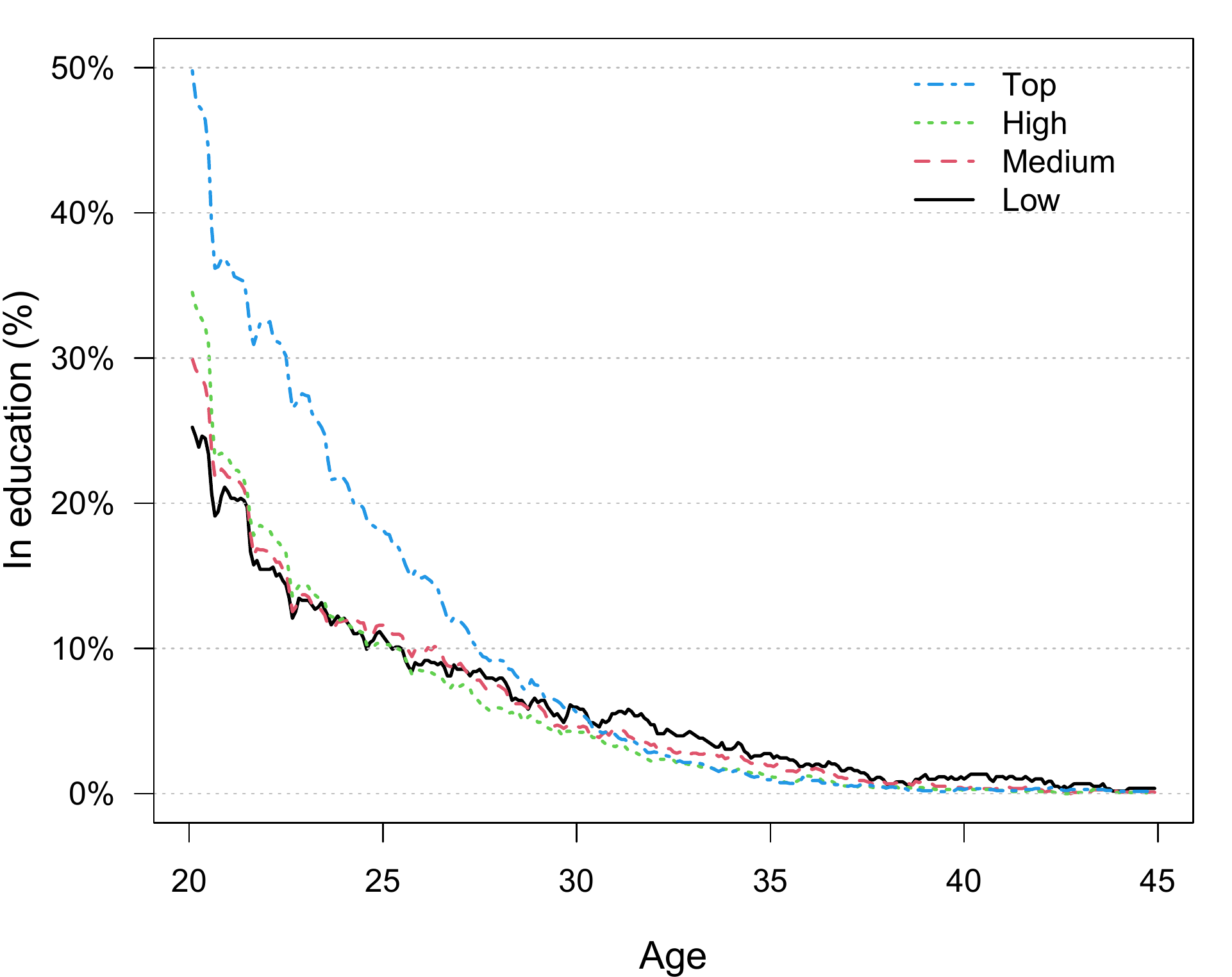}
\caption{\label{EarningsByEduc:Fig}Age-earning trajectories of
  nulliparous women by highest earnings at age 35-44, cohorts 1950-74,
  West Germany.}
\end{figure}

It is clear that the earning
trajectories of women greatly differ, depending on the career
track. In order to illustrate the differences in women's earning
profiles, Figure \ref{EarningsByEduc:Fig} plots the average earnings
(left panel) and the share of women in education (right panel) by
their highest earnings at age 35-44.  The figure shows that the
age-earning profiles of the women who eventually reach high earnings
are flat at early ages, but fairly steep at advanced ages.
Differences in earnings at later ages can
largely be attributed to differences in educational participation (right
panel). This aspect is of relevance, when
prototypical trajectories for monthly earnings and
employment status will be used below to illustrate the model results.

\begin{figure}\centering
\includegraphics[width=1.\textwidth]{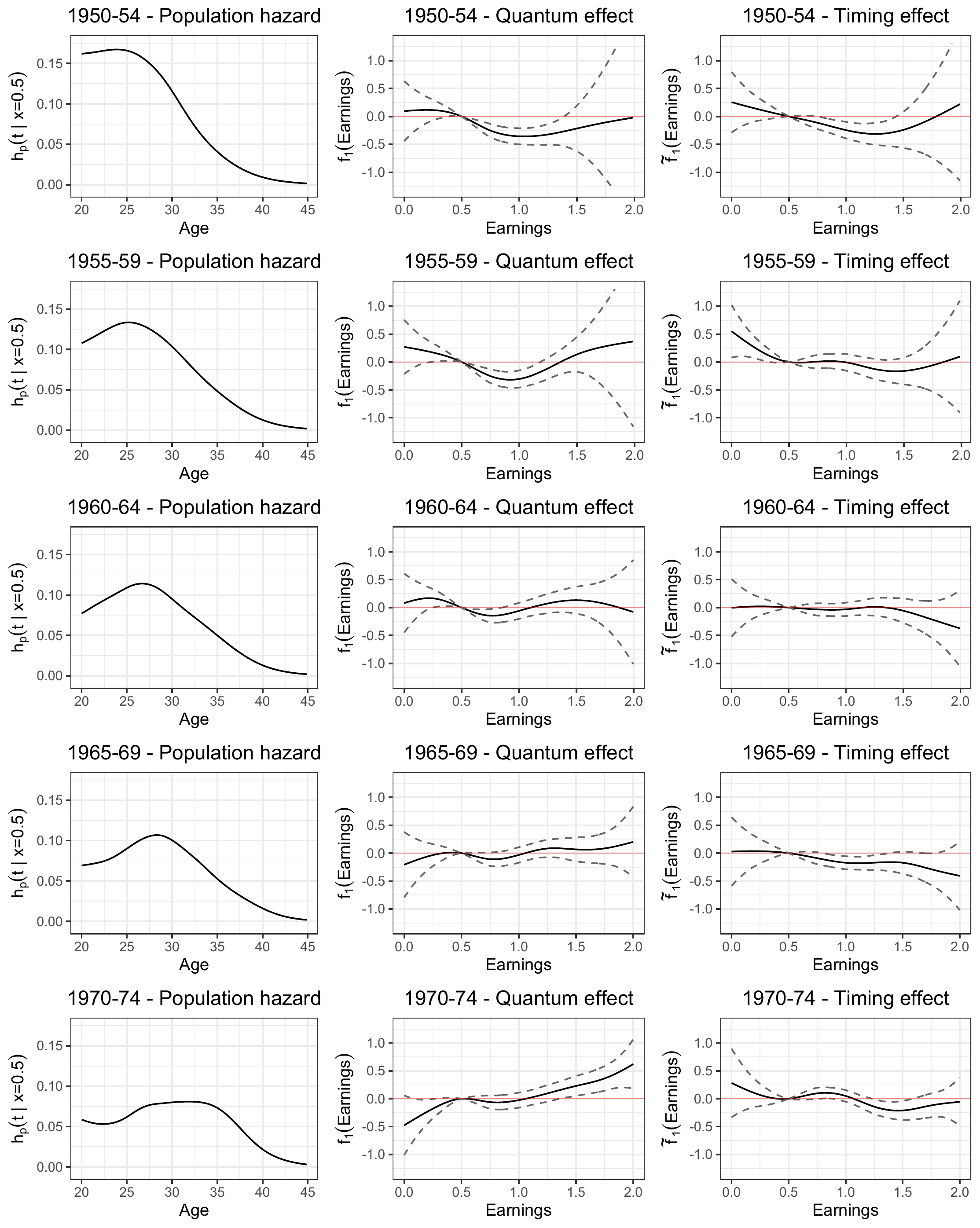}
\caption{\label{App:AdditiveTerms:Fig} Estimated reference hazard
  $\mathrm{e}^{\beta_0}f_0(t)$ and earnings effects on the quantum and
  timing of a first pregnancy with earnings reference value set at
  one-half of its annual average value in model $\mathcal{M}_1$.}
\end{figure}

Analyses based on the TVcure model described in Section
\ref{TVcure:Sec} were made separately
for the groups of birth cohorts mentioned above.
Seven TVcure models were fitted to the data,
see Table \ref{App:FittedModels:Tab}
for the included covariates
in the additive submodels for long-term (Quantum) and short-term
(Timing) survival. For example, model $\mathcal{M}_1$ assumes joint
effects of the employment status and earnings both on event quantum
and timing.
\begin{table}
  \caption{\label{App:FittedModels:Tab}TVcure models fitted to the West German women's cohort
    data: \ttf{Status} refers to the categorical covariate indicating the employment
    status and \ttf{s(Earnings)} to an additive earnings effect on
      long-term (Quantum) and/or short-term (Timing) survival.}\mbox{}
  \centering
    \begin{tabular}{llccccccc}
\toprule
  & & \multicolumn{7}{c}{Model} \\
       \cmidrule(lr){3-9}
      Submodel & Covariates  & $\mathcal{M}_1$ & $\mathcal{M}_2$ & $\mathcal{M}_3$
                             & $\mathcal{M}_4$ & $\mathcal{M}_5$ & $\mathcal{M}_6$ & $\mathcal{M}_7$ \\
      \hline
   Quantum & \ttf{Status}        & \chck & \chck & \chck & \chck & \chck & --   & -- \\
           & \ttf{s(Earnings)} & \chck & \chck &  --   &  --   & --      & --   & -- \\
   Timing  & \ttf{Status}        & \chck & \chck & \chck & \chck & --    & \chck& -- \\
           & \ttf{s(Earnings)} & \chck &   --  & \chck &  --   & --      & --   & -- \\
\bottomrule
    \end{tabular}
\end{table}
The model fit measured by the deviance,
$D=2\sum_{i=1}^n\sum_{j=1}^{n_i}\left\{
  d_{ij}\log(d_{ij}/\mu_{ij})-(d_{ij}-\mu_{ij}) \right\}$, and the
model complexity quantified by the effective degrees of freedom (EDF)
given by the trace of $\mathbf{H}_{\tau}^{-1}\mathbf{H}_{0}$
\citep{HastieTibshirani1990}, can be combined to obtain the Akaike
Information Criterion, $\text{AIC}=D+2\,\text{EDF}$.  A model
selection relying on the AIC, see Table \ref{App:AIC:Tab}, suggests a
significant effect of employment status (working, studying, unemployed
or other reasons for not being in the labour market)
on the probability to have a first child and on the timing of the
pregnancy for all cohorts.
There is also a statistically significant association between female
earnings and first birth timing and quantum, with the exception of the
1960-64-cohorts. For the latter cohorts, earnings do not affect
fertility tempo. 
\begin{table}
    \caption{\label{App:AIC:Tab} Deviance, effective degrees of
      freedom (EDF) and Akaike Information Criterion (AIC) for the
      TVcure models in Table \ref{App:FittedModels:Tab}} 
    \centering
  \begin{tabular}{crrrcrrr}
\toprule
    & \multicolumn{3}{c}{Cohort 1950-54}
    && \multicolumn{3}{c}{Cohort 1955-59}  \\
    \cmidrule(lr){2-4} \cmidrule(lr){6-8} 
  Model & Deviance & EDF & AIC && Deviance & EDF & AIC \\ 
  \hline
$\mathcal{M}_1$ & 18055.72 & 14.2 & \ul{18084.19} && 17460.17 & 14.8 & \ul{17489.84} \\ 
$\mathcal{M}_2$ & 18069.87 & 11.0 & 18091.81 && 17469.91 & 11.1 & 17492.07 \\ 
$\mathcal{M}_3$ & 18082.72 & 10.9 & 18104.45 && 17491.76 & 11.0 & 17513.82 \\ 
$\mathcal{M}_4$ & 18125.19 &  7.0 & 18139.19 && 17518.86 &  7.0 & 17532.86 \\ 
$\mathcal{M}_5$ & 18134.80 &  4.0 & 18142.80 && 17528.57 &  4.0 & 17536.57 \\ 
$\mathcal{M}_6$ & 18145.99 &  4.0 & 18153.99 && 17581.49 &  4.0 & 17589.49 \\ 
$\mathcal{M}_7$ & 18209.75 &  1.0 & 18211.75 && 17644.05 &  1.0 & 17646.05 \\
    \hline
\\
    & \multicolumn{3}{c}{Cohort 1960-64}
    && \multicolumn{3}{c}{Cohort 1965-69}  \\
    \cmidrule(lr){2-4} \cmidrule(lr){6-8} 
  Model & Deviance & EDF & AIC && Deviance & EDF & AIC \\ 
  \hline
$\mathcal{M}_1$ & 22247.38 & 15.4 & 22278.24 && 25007.62 & 15.7 & \ul{25038.94}\\ 
$\mathcal{M}_2$ & 22250.92 & 11.4 & \ul{22273.80} && 25019.54 & 11.7 & 25042.88\\ 
$\mathcal{M}_3$ & 22263.40 & 11.3 & 22285.94 && 25017.12 & 11.3 & 25039.80\\ 
$\mathcal{M}_4$ & 22274.21 &  7.0 & 22288.21 && 25034.64 &  7.0 & 25048.64\\ 
$\mathcal{M}_5$ & 22288.15 &  4.0 & 22296.15 && 25045.19 &  4.0 & 25053.19\\ 
$\mathcal{M}_6$ & 22347.23 &  4.0 & 22355.23 && 25093.10 &  4.0 & 25101.10\\ 
$\mathcal{M}_7$ & 22491.47 &  1.0 & 22493.47 && 25288.45 &  1.0 & 25290.45\\
    \hline \\
    & \multicolumn{3}{c}{Cohort 1970-74} \\
    \cmidrule(lr){2-4}
  Model & Deviance & EDF & AIC \\
  \cline{1-4}
$\mathcal{M}_1$ & 29357.81 & 16.4 & \ul{29390.52} \\ 
$\mathcal{M}_2$ & 29374.80 & 11.8 & 29398.33 \\ 
$\mathcal{M}_3$ & 29387.04 & 11.5 & 29410.10 \\ 
$\mathcal{M}_4$ & 29404.60 &  7.0 & 29418.60 \\ 
$\mathcal{M}_5$ & 29436.07 &  4.0 & 29444.07 \\ 
$\mathcal{M}_6$ & 29468.11 &  4.0 & 29476.11 \\ 
$\mathcal{M}_7$ & 29660.88 &  1.0 & 29662.88 \\
  \cline{1-4}
\end{tabular}
\end{table}
Figure \ref{App:AdditiveTerms:Fig} plots the estimates for model
$\mathcal{M}_1$ separately for each cohort. The first row displays the
first birth pattern of a woman with half the average gross
earnings. It shows clearly how the fertility schedule has shifted
across birth cohorts. First birth has been postponed, at the same time
the distribution has become wider, suggesting greater heterogeneity in
the age at first parenthood. The estimates of the additive terms for
earnings in the quantum and timing are displayed in the 2nd and 3rd
columns of Fig.\,\ref{App:AdditiveTerms:Fig}.
These results suggest that employed West German women in
the 1950's who had low earnings were more likely to become mothers
at a younger age than those who earned more. The association flips
across cohorts, with the 1960-64 cohort playing a pivotal role.
For the recent cohorts, the association is now positive, with low earnings
reducing the chances of having a first child. The patterns for the
younger cohorts (1965-69 and 1970-74) are very similar,
with an increasing influence of earnings on the decision to have a
first child, with some delay for the better-off.

An approach based on TVcure models
with a categorized version of monthly earnings, \ttf{EarnCat} (for
classification, see above), was also explored for robustness
checks. Four TVCure models with or without \ttf{EarnCat} in the
quantum or the timing submodels 
were fitted separately for each
cohort of women, with deviance, EDF and AIC
also computed.
Deviances (not reported here to save space)
are, unsurprisingly, larger than those obtained with the continuous
version of earnings given the loss of information resulting from
categorization.
However, qualitative conclusions are coherent
with significant joint effects of \ttf{EarnCat} on the probability to
have a first child and on the timing of the pregnancy for all cohorts.
The analysis shows that the economic prerequisite for having children
have shifted for the recent cohorts. While periods of low earnings increased
``fertility quantum'' for the older cohorts, it is rather vice versa
for the younger cohorts.
\begin{figure}\centering
  \begin{tabular}{c}
    {Differences by earning trajectory within cohort groups}\\
    \includegraphics[width=\textwidth]{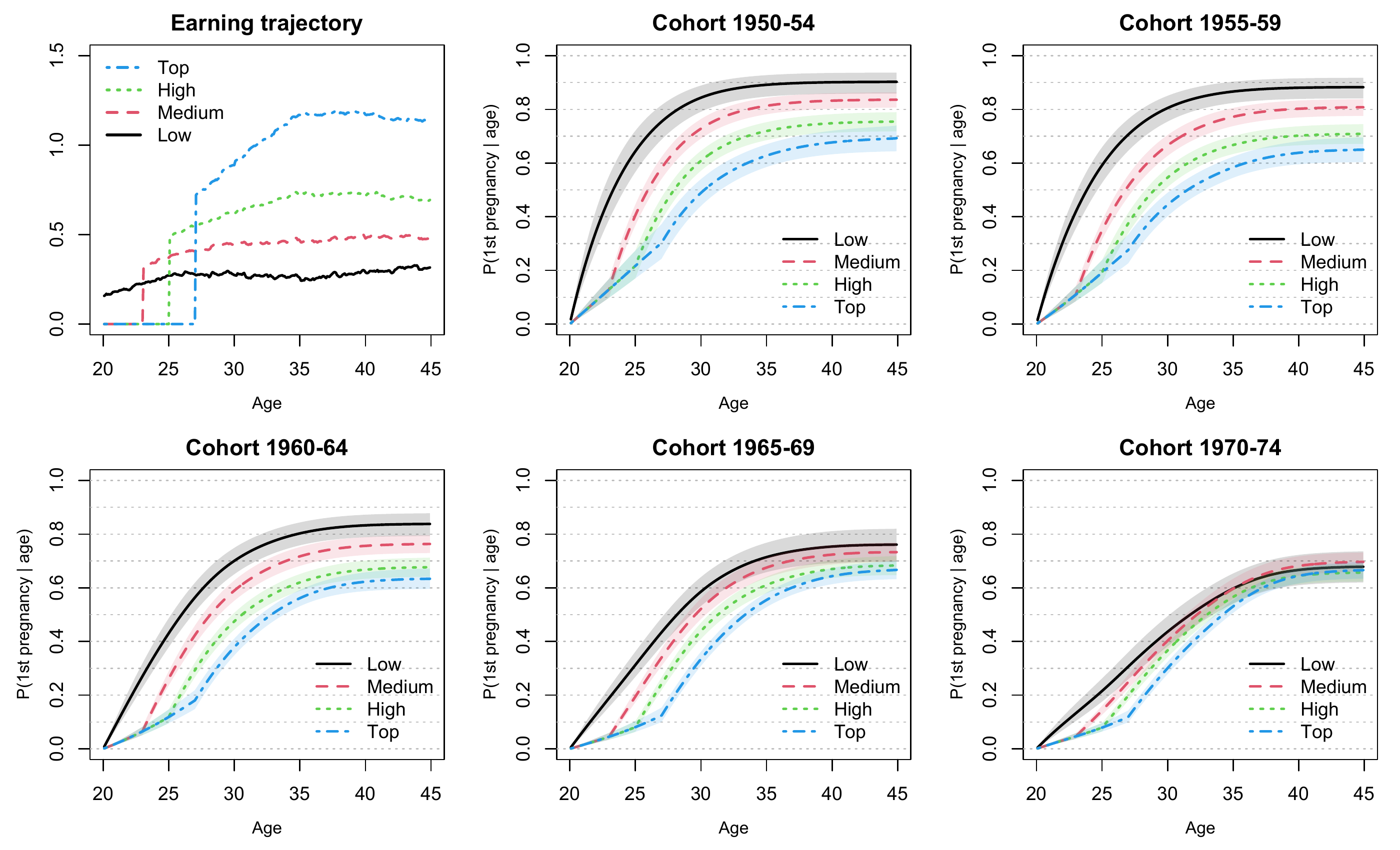} \\
    {Differences by cohorts within earning trajectory groups}\\
    \includegraphics[width=\textwidth]{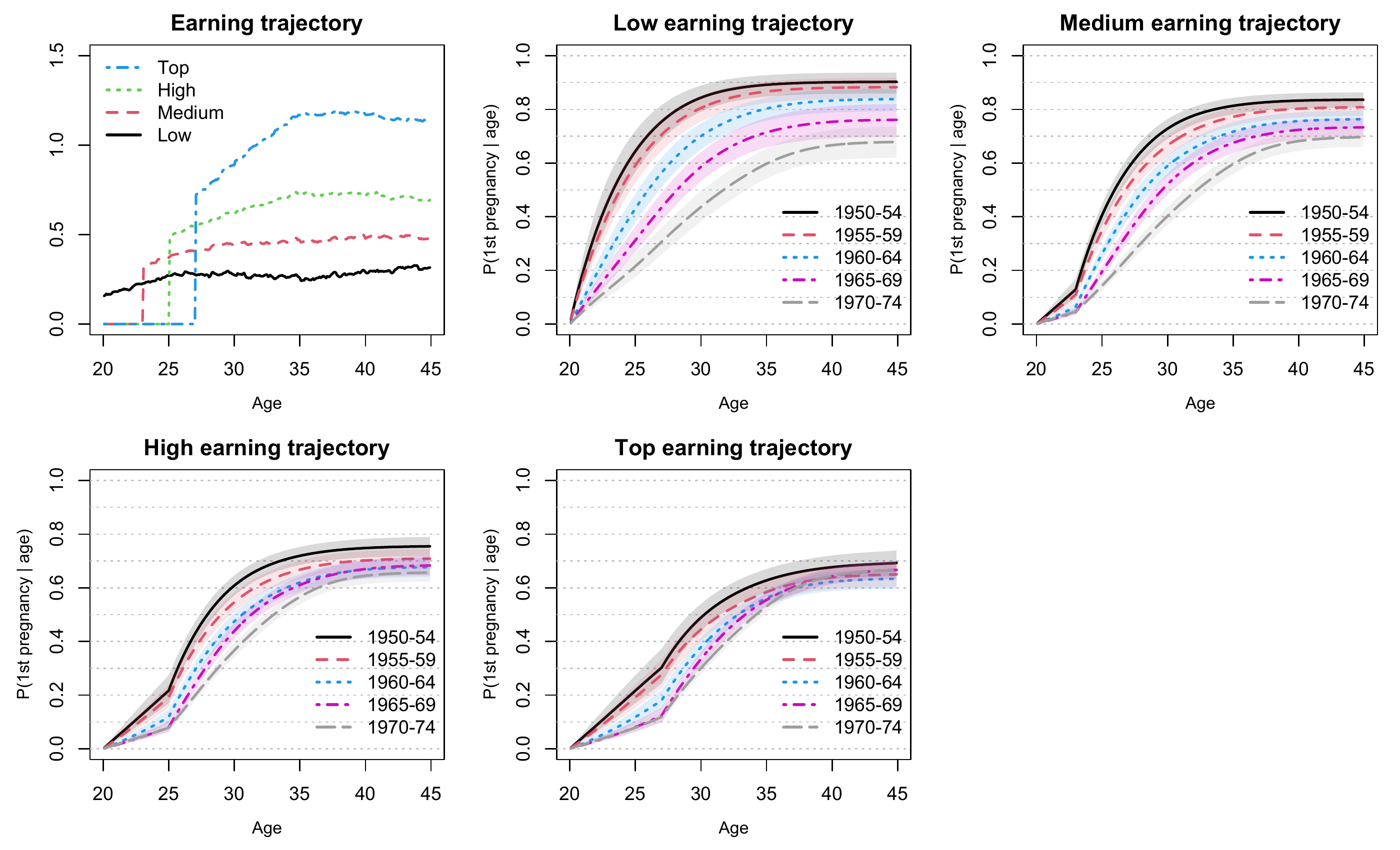}\\
  \end{tabular}                                                                
  \caption{\label{App:FittedFp:Fig} Estimated probability of a first
    pregnancy of West German women by prototypical earning profiles
    and birth cohort.}
\end{figure}

So far, the analysis has focused on the association between earnings
and first birth behavior. We have ignored that some women may have low
incomes at an early age, while they may earn high wages later in
life. To account for that, we estimated the conditional probability of
being pregnant for four `prototypical' earning trajectories using the
fitted TVCure models. We have selected the following four scenarios,
its descriptive name referring to the highest earning value at age
35-44 (see also the top left graph of Fig.\,\ref{App:FittedFp:Fig}):
\begin{enumerate}
  \setlength{\itemsep}{0pt}
  \item `Low': a woman with low earning trajectory studying until age 20 (solid
    line);
  \item `Medium': a woman with medium earning trajectory after studying until age
    22 (dashed line);
  \item `High': a woman who completes her education at age 25 and then
    moves to the high earning trajectory (dotted line);
  \item `Top': A woman who completes her education at age 27 and moves
    to the top earning trajectory (dashed-dotted line).
\end{enumerate}  
We have displayed the estimates by birth cohort and earning group
(top and bottom panels of Fig.\,\ref{App:FittedFp:Fig},
respectively).  It shows how the association of women's earning and
first birth progression has shifted across birth cohorts. While there
were large fertility differences by women's earning profiles for the
older cohorts, patterns have become more similar across time. We have
also flipped the figure around and displayed the developments across
cohorts within earning groups (lower panel of
Fig.\,\ref{App:FittedFp:Fig}). While the fertility schedule is
relatively stable across cohorts for high and top earning
trajectories, first birth is increasingly postponed for low- and
medium-income women.

\section{Discussion} \label{Discussion:Sec}
The proposed methodology extends cure survival models by enabling the
inclusion of time-varying covariates in the quantum and timing
additive submodels.  They are not restricted to the conditional
survival model for non-cured subjects as in \citet{Dirick2019}, nor
are the number of changes in covariate values limited as in
\citet{LambertBremhorst2020} where only a small number of
exclusively categorical covariate updates were allowed.  Our new
specification enables the inclusion of time-varying quantitative
explanatory variables such as frequently changing monthly earnings in
the fertility application. Smooth nonlinear forms can also be assumed
for these effects with values for the penalty parameter $\pmb{\tau}$
associated to P-splines automatically selected using Laplace-based
approximations to the marginal posterior distributions of
$\pmb{\tau}$.

In a last step of the investigation, we used register data for Germany
to illustrate how the method worked with real data. We examined
whether the effect of women's earnings on first birth behavior had
changed across cohorts in Germany. Data and research question seemed
an ideal test case to showcase how the method unfolds in
practice. There is a high share of women who remain childless in
Germany. Thus, there is a sizeable cure fraction for each
cohort. Further, Germany has enacted major policy reforms in the last
decades. Particularly the parental leave benefit reform, enacted in
2007, set stronger incentives than before to postpone childbearing
until one had reached ``sufficient'' earnings. It was
expected that 
it would lower childlessness among the highly educated and
career-oriented women in Germany. Our investigation, that separated
timing and quantum, provides important and social policy relevant
evidence on the matter.
We indeed find that high earnings used to increase childlessness in
the old cohorts, while we no longer find such a relationship for the
younger cohorts.
The analysis shows that the association between women's earnings and
first birth has changed across birth cohorts. While low earnings used
to accelerate fertility timing and reduce childlessness among the
older cohorts, we do not find the same patterns anymore for the recent
cohorts. Indeed, levels of childlessness are very similar now,
regardless of earnings. Insufficient female earnings seem to
increasingly delay childbearing in Germany. This pattern may be
attributed to the policy reform and the introduction of the
earnings-related parental leave benefit which set strong incentives to
postpone parenthood until one had gathered earnings that resulted into
adequate parental leave benefits.  

\section*{Acknowledgments}
The first author acknowledges the support of the ARC project IMAL
(grant 20/25-107) financed by the Wallonia-Brussels Federation and
granted by the Acad\'emie universitaire Louvain.

\bibliography{LambertKreyenfeld_arXiv1.bib}

\appendix
\section*{Appendices}

\section{Closed form expressions for 
  $\mathbf{U}_\lambda$ and $\mathbf{H}_\lambda$}\label{Appendix:A}
Let $b_{tk}=b_k(t)$, $\pi_t=f_0(t)\,dt=
{\exp\big(\sum_{k=1}^K b_{tk}\phi_k\big)/ \sum_{s=1}^T \exp\big(\sum_{k=1}^K b_{sk}\phi_k\big)}$,
$h_{it}=h_p(t|\mathbf{v}_i(t),\tilde{\mathbf{v}}_i(t))$, $F_0(t)=\sum_{s\leq t}\pi_t$ and $S_0(t)=1-F_0(t)$.
For the spline parameters involved in $f_0(t)$, the conditional score and precision matrix are
\begin{align}
  \begin{split}
  \mathbf{U}_{\tau_0}^\phi
  &= {\partial \ell_p \over \partial \pmb{\phi}}
  = \sum_{i=1}^n \sum_{t=1}^{t_i}(d_{it}-\mu_{it}) {\partial \log h_{it} \over \partial \pmb{\phi}}
    -\tau_0 \mathbf{P}\pmb{\phi} \\
  -\mathbf{H}_{\tau_0}^{\phi\phi}
  &=
    \sum_{i=1}^n \sum_{t=1}^{t_i}\mu_{it} {\partial \log h_{it} \over \partial \pmb{\phi}}
    {\partial \log h_{it} \over \partial \pmb{\phi}^\T}
    +\tau_0 \mathbf{P}
  \end{split}
        \label{GradHesPhi:Eq}
\end{align}
with
\begin{align*}
&  {\partial \log h_{it} \over \partial \phi_k}
  =
    {\partial \log f_0(t) \over \phi_k} +
    (\ee{\eta_F(\tilde{\mathbf{v}}_i(t))}-1) {\partial \log S_0(t) \over \phi_k} \\
&  {\partial \log f_0(t) \over \phi_k} 
  = \breve{b}_{tk} = b_{tk} - \sum_s\pi_s b_{sk} \\
&  {\partial \log S_0(t) \over \phi_k}
  = -{1 \over S_0(t)}\sum_{s\leq t} \pi_s \breve{b}_{sk} ~.
\end{align*}
For the regression and spline parameters defining long-term
survival, the conditional score and precision matrix are
\begin{align}
  \begin{split}
  \mathbf{U}_{\lambda}^\psi
  &= {\partial \ell_p \over \partial \pmb{\psi}}
    = \sum_{i=1}^n \sum_{t=1}^{t_i}(d_{it}-\mu_{it}) {\partial \log h_{it} \over \partial \pmb{\psi}}
    -\mathbf{K}_\lambda ({\pmb{\psi}}-\mathbf{b}) \\
  -\mathbf{H}_{\lambda}^{\psi\psi}
  &=
    \sum_{i=1}^n \sum_{t=1}^{t_i}\mu_{it} {\partial \log h_{it} \over \partial \pmb{\psi}}
    {\partial \log h_{it} \over \partial \pmb{\psi}^\T}
    +\mathbf{K}_\lambda
  \end{split}
    \label{GradHessPsi:Eq}
\end{align}
with
${\partial \log h_{it} / \partial \psi_k} = ({\mathbfcal X}_t)_{ik}$.
For the regression and spline parameters defining short-term
survival, the conditional score and precision matrix are
\begin{align}
  \begin{split}
  \mathbf{U}_{\tilde{\lambda}}^{\tilde{\psi}}
  &= {\partial \ell_p \over \partial \tilde{\pmb{\psi}}}
    = \sum_{i=1}^n \sum_{t=1}^{t_i}(d_{it}-\mu_{it}) {\partial \log h_{it} \over \partial \tilde{\pmb{\psi}}}
    -\tilde{\mathbf{K}}_{\tilde{\lambda}} ({\tilde{\pmb{\psi}}}-\mathbf{g}) \\
  -\mathbf{H}_{\tilde{\lambda}}^{\tilde{\psi}\tilde{\psi}}
  &=
    \sum_{i=1}^n \sum_{t=1}^{t_i}\mu_{it} {\partial \log h_{it} \over \partial \tilde{\pmb{\psi}}}
    {\partial \log h_{it} \over \partial \tilde{\pmb{\psi}}^\T}
    +\tilde{\mathbf{K}}_{\tilde{\lambda}}
  \end{split}
    \label{GradHesPsiTilde:Eq}
\end{align}
with
${\partial \log h_{it} / \partial \tilde{\psi}_k} =
({\tilde{\mathbfcal X}}_t)_{ik}
\big(
   1 + \ee{\eta_F(\tilde{\mathbf{v}}(t))} \log S_0(t)
\big)
$.
Closed forms for cross-derivatives (independent of the penalty parameters) can also be obtained:
\begin{align*}
  -\mathbf{H}^{\psi\tilde{\psi}}
  &=
    \sum_{i=1}^n \sum_{t=1}^{t_i}\mu_{it} {\partial \log h_{it} \over \partial \pmb{\psi}}
    {\partial \log h_{it} \over \partial \tilde{\pmb{\psi}}^\T} \,.
\end{align*}
Let $\pmb{\zeta}=(\pmb{\psi}^\T,\tilde{\pmb{\psi}}^\T)^\T$ and
$\pmb{\tau}=(\lambda^\T,\tilde{\lambda}^\T)^\T$.
Then, the score and (minus) the precision matrix for $\pmb{\zeta}$ are
\begin{align} 
  \mathbf{U}^\zeta_\tau=
\begin{pmatrix}
  \mathbf{U}^\psi_\lambda \\[.3em]
  \mathbf{U}^{\tilde{\psi}}_{\tilde{\lambda}}
\end{pmatrix}  
  ~~;~~
\mathcal{H}_\tau =
\begin{bmatrix}
  \mathbf{H}_{\lambda}^{\psi\psi} & \mathbf{H}^{\psi\tilde{\psi}} \\
  \mathbf{H}^{\tilde{\psi}\psi}   & \mathbf{H}_{\tilde{\lambda}}^{\tilde{\psi}\tilde{\psi}}
\end{bmatrix} ~.
 \label{GradHessZeta:Eq}
\end{align}
  
\end{document}